\documentclass[preprint]{aastex}

\begin{document}

\title{Optical and Infrared Spectroscopy of SN~1999ee and SN~1999ex \altaffilmark{1}}
\author{Mario Hamuy\altaffilmark{2,3} }
\affil{The Observatories of the Carnegie Institution of Washington, 813 Santa Barbara Street, Pasadena, CA 91101}
\email{mhamuy@ociw.edu}
\author{Jos\'{e} Maza\altaffilmark{3} \altaffilmark{4} }
\affil{Departamento de Astronom\'\i a, Universidad de Chile, Casilla 36-D, Santiago, Chile}
\email{jose@das.uchile.cl}
\author{Philip A. Pinto}
\affil{Steward Observatory, The University of Arizona, Tucson, AZ 85721}
\email{ppinto@as.arizona.edu}
\author{M. M. Phillips}
\affil{Carnegie Institution of Washington, Las Campanas Observatory, Casilla 601, La Serena, Chile} 
\email{mmp@lco.cl}
\author{Nicholas B. Suntzeff}
\affil{National Optical Astronomy Observatories\altaffilmark{5}, Cerro Tololo Inter-American Observatory, Casilla 603, La Serena, Chile}
\email{nsuntzeff@noao.edu}
\author{R. D. Blum}
\affil{National Optical Astronomy Observatories\altaffilmark{5}, Cerro Tololo Inter-American Observatory, Casilla 603, La Serena, Chile}
\email{rblum@noao.edu}
\author{K.A.G. Olsen}
\affil{National Optical Astronomy Observatories\altaffilmark{5}, Cerro Tololo Inter-American Observatory, Casilla 603, La Serena, Chile}
\email{kolsen@noao.edu}
\author{David J. Pinfield\altaffilmark{3}}
\affil{Astrophysics Research Institute, Liverpool John Moores University, Twelve Quays House, Egerton Wharf, Birkenhead, CH41 1LD, UK}
\email{dpi@astro.livjm.ac.uk}
\author{Valentin D. Ivanov}
\affil{European Southern Observatory, Alonso de C\'ordova 3107, Vitacura, Casilla 19001, Santiago 19, Chile}
\email{vivanov@eso.org}       
\author{T. Augusteijn\altaffilmark{6}}
\affil{European Southern Observatory, Alonso de C\'ordova 3107, Vitacura, Casilla 19001, Santiago 19, Chile}
\email{tau@ing.iac.es}
\author{S. Brillant}
\affil{European Southern Observatory, Alonso de C\'ordova 3107, Vitacura, Casilla 19001, Santiago 19, Chile}
\email{sbrillan@eso.org}
\author{M. Chadid}
\affil{European Southern Observatory, Alonso de C\'ordova 3107, Vitacura, Casilla 19001, Santiago 19, Chile}
\email{mchadid@eso.org}
\author{J.-G. Cuby}
\affil{European Southern Observatory, Alonso de C\'ordova 3107, Vitacura, Casilla 19001, Santiago 19, Chile}
\email{jcuby@eso.org} 
\author{V. Doublier}
\affil{European Southern Observatory, Alonso de C\'ordova 3107, Vitacura, Casilla 19001, Santiago 19, Chile}
\email{vdoublie@eso.org}
\author{O. R. Hainaut}
\affil{European Southern Observatory, Alonso de C\'ordova 3107, Vitacura, Casilla 19001, Santiago 19, Chile}
\email{ohainaut@eso.org}
\author{E. Le Floc'h}
\affil{European Southern Observatory, Alonso de C\'ordova 3107, Vitacura, Casilla 19001, Santiago 19, Chile}
\email{elefloch@eso.org}
\author{C. Lidman}
\affil{European Southern Observatory, Alonso de C\'ordova 3107, Vitacura, Casilla 19001, Santiago 19, Chile}
\email{clidman@eso.org}
\author{Monika G. Petr-Gotzens\altaffilmark{7}}
\affil{European Southern Observatory, Alonso de C\'ordova 3107, Vitacura, Casilla 19001, Santiago 19, Chile}
\email{mpetr@mpifr-bonn.mpg.de}
\author{E. Pompei}
\affil{European Southern Observatory, Alonso de C\'ordova 3107, Vitacura, Casilla 19001, Santiago 19, Chile}
\email{epompei@eso.org}
\author{L. Vanzi}
\affil{European Southern Observatory, Alonso de C\'ordova 3107, Vitacura, Casilla 19001, Santiago 19, Chile}
\email{lvanzi@eso.org} 

\altaffiltext{1}{Based on observations collected at the European Southern Observatory, Chile (program ESO 164.H-0376).} 
\altaffiltext{2}{Hubble Fellow}
\altaffiltext{3}{Visiting Astronomer, European Southern Observatory.}
\altaffiltext{4}{Visiting Astronomer, Cerro Tololo Inter-American Observatory.
CTIO is operated by AURA, Inc. under contract to the National Science Foundation.}
\altaffiltext{5}{Cerro Tololo Inter-American Observatory, Kitt Peak
National Observatory, National Optical Astronomy Observatories,
operated by the Association of Universities for Research in Astronomy,
Inc., (AURA), under cooperative agreement with the National Science
Foundation.}
\altaffiltext{6}{Present address: Isaac Newton Group of Telescopes, Apartado de Correos 321, 38700 Santa Cruz de La Palma, Canary Islands, Spain}
\altaffiltext{7}{Present address: Max-Planck-Institut f\"ur Radioastronomie, Auf dem H\"ugel 69, D-53121 Bonn, Germany}

\begin{abstract}

We report optical and infrared spectroscopic observations of the Type Ia SN~1999ee
and the Type Ib/c SN~1999ex, both of which were hosted by the galaxy IC~5179.
For SN~1999ee we obtained a continuous sequence with an unprecedented
wavelength and temporal coverage beginning 9 days before maximum light and
extending through day 42.  Before maximum light SN~1999ee displayed a normal
spectrum with a strong Si II $\lambda$6355 absorption, thus showing that not
all slow-declining SNe are spectroscopically peculiar at these evolutionary phases. 
A comparative study of the infrared spectra of SN~1999ee and other Type Ia supernovae shows 
that there is a remarkable homogeneity among the Branch-normal SNe Ia during their
first 60 days of evolution. SN~1991bg-like objects, on the other hand, display
spectroscopic  peculiarities at IR wavelengths.
SN~1999ex was characterized by  the lack of hydrogen lines,
weak optical He I lines, and strong He I $\lambda$$\lambda$10830,20581, thus
providing an example of an intermediate case between pure Ib and Ic supernovae.
We conclude therefore that SN~1999ex provides first clear evidence
for a link between the Ib and Ic classes and that there is a continuous 
spectroscopic sequence ranging from the He deficient SNe~Ic to the SNe~Ib 
which are characterized by strong optical He I lines.

\end{abstract}


\section{Introduction}

The last ten years have witnessed an enormous progress in our
knowledge of the optical properties of supernovae (SNe) of all types.
However, comparatively little is still known about these objects in
near-infrared (NIR) wavelengths.
Given the rapid technological development of NIR light detection
over recent years, in 1999 we started a program to
obtain optical and NIR photometry and spectroscopy of nearby SNe ($z$$<$0.08).
So far, the ``Supernova Optical and Infrared Survey'' (SOIRS)
program has gathered high-quality observations for 
$\sim$20 SNe. So far we have reported results for the bright Type II SN~1999em \citep{hamuy01}.
In this paper we report spectroscopic observations of two of the best-observed objects included in our program, the Type Ia
SN~1999ee and the Type Ib/c SN~1999ex, both of which exploded in the same galaxy within three weeks.

SN~1999ee was discovered  by M. Wischnjewsky on a film taken on 1999 October 7.15 (JD 2451458.65) 
in the course of the 
El Roble survey \citep{maza99}. The SN exploded 10 arcsec East and 10 arcsec South
of the nucleus of the spiral galaxy IC 5179, a very active star-forming galaxy with a heliocentric
redshift of 3,498 km~s$^{-1}$.
The optical spectrum taken on 1999 October 9.10 revealed the defining Si II $\lambda$6355 feature
of the Type Ia class \citep{maza99}. The high expansion velocity of 15,700 km~s$^{-1}$ 
deduced from this line and the faint apparent magnitude at discovery ($\sim$17.5)
indicated that this object had been found several days before maximum light.
At the distance of the host galaxy SN~1999ee offered the promise to reach
$V$=14.5 about 10 days later, thus proving to be an excellent
target for a detailed study of a SN Ia early since explosion, both at optical and IR wavelengths.
This discovery occurred at the very beginning of one of the
SOIRS follow-up runs, previously scheduled for 1999 October-November,
so we decided to give first priority to SN~1999ee.
The unique opportunity afforded by SN~1999ee led the ESO Director General
to allocate director's discretionary time to this project in order
to secure the best possible data for SN~1999ee. As a result of this effort
we obtained a superb dataset of optical/IR photometric and spectroscopic
observations of the first 50 days of the evolution of SN~1999ee. 

Three weeks after the discovery of SN~1999ee, a second SN exploded in the same galaxy that hosted SN~1999ee.
The discovery of SN~1999ex was made by Martin et al. (1999) on 1999 November 9.51 UT (JD 2451492.01)
in the course of the PARG Automated Supernova Search at Perth Observatory.
They reported that the SN was not visible in a deep exposure taken on 1999 October 25.58
(estimated limiting mag 19), and that the object was slowly brightening.
A spectrum taken on 1999 November 14.13 \citep{hamuy99} showed that SN~1999ex had
close resemblance to that of the Type Ic SN 1994I taken near maximum light
\citep{filippenko95}. This led us initially to classify SN~1999ex as a Ic
event although we believe that it should be typed as an intermediate Ib/c
object (see Sec. \ref{sn99ex.sec}). As soon as this object was discovered we decided
to include it in our optical/IR spectroscopic follow-up.

The observations gathered for SNe~1999ee and 1999ex have an unprecedented
temporal and  wavelength coverage and afford the possibility to carry out a detailed comparison
with atmosphere models.  In this paper we report 
the spectroscopic observations of SNe~1999ee and 1999ex and
we discuss these results. The spectra presented here are available 
in electronic form to other researchers (contact M. H. if interested).
Optical and IR photometry will be published and discussed
in detail elsewhere by \citet{stritzinger02} and \citet{krisciunas02}, respectively.

\section{Spectroscopic Observations and Reductions}

We obtained optical and IR spectra of SNe~1999ee and 1999ex with 
the ESO NTT/EMMI/SOFI, the Danish 1.5-m/DFOSC equipments at La Silla,
and the VLT/ISAAC instrument at Cerro Paranal between 1999 October 9
and 1999 November 28. We also obtained three optical spectra with the CTIO~4-m telescope
on 1999 October 9, the CTIO~1.5-m telescope on 1999 October 27,
and the Las Campanas Dupont 2.5-m telescope on 1999 October 16.
Table 1 gives the journal of the observations.

\subsection{Optical Spectroscopy}

The NTT observations comprised three different setups.
We used the blue channel of EMMI equipped 
with a Tek CCD (1024x1024) and grating 5 (158 lines mm$^{-1}$) which,
in first order, delivered spectra with a dispersion of 3.5~\AA~pix$^{-1}$ and a useful
wavelength range between 0.330 and 0.525 $\mu m$. With the red channel,
CCD Tek 2048, and grating 13 (150 lines mm$^{-1}$) the dispersion was 2.7~\AA~pix$^{-1}$
and the spectral coverage included from 0.470 through 1.100 $\mu m$ in first order.
Since this setup had potential second-order contamination beyond $\sim$0.6 $\mu m$
we decided to take one spectrum with the OG530 filter and a second observation
without the filter, in order to provide an overlap with the blue spectrum.
Thus, a single-epoch observation comprised three spectra.

The observations with EMMI started with calibrations during day time 
(bias and dome flat-field exposures). The night began
with the observation of a spectrophotometric standard (from
the list of \citet{hamuy94}) through a wide slit of 10 arcsec, after which
we observed the SNe with a slit of 1 arcsec. Before the discovery of
SN~1999ex on 1999 November 9 we oriented the slit along the line connecting
SN~1999ee and the host galaxy nucleus. On 1999 November 14 and 19, on the other hand,
we oriented the slit along the two SNe in order to get simultaneous
spectra. Since the airmass of our observations was always below 1.5
we do not expect serious systematic errors in the observed fluxes
due to atmospheric refraction.
We took two exposures of the SNe per spectral setup, each of the same length  
(typically 300-600 sec). Immediately following this observation we observed
a He-Ar lamp, at the same position of the SNe and before
changing the optical setup in order to ensure an accurate wavelength
calibration.  At the end of the night we observed a second flux standard.

We also used the Danish 1.5-m telescope and the DFOSC instrument at La Silla 
on five nights, between 1999 October 20 and November 28, in order to improve the 
temporal coverage of our spectroscopic observations. In all cases
we employed a 2K$\times$2K LORAL CCD and two different grisms to secure
a spectral coverage comparable to that obtained with EMMI.
With grism 3 (400 lines mm$^{-1}$) we covered a useful wavelength
range between 0.33 and 0.66 $\mu m$ with a dispersion of 2.3~\AA~pix$^{-1}$.
With grism 5 (300 lines mm$^{-1}$) we sampled the range between
0.53 and 0.98 $\mu m$ at 3.1~\AA~pix$^{-1}$. The red setup produced
spectra with significant fringing beyond 0.75 $\mu m$ which we
did not attempt to remove. 
We observed the SNe with a slit of 2 arcsec. Although we did not
rotate the slit along the parallactic angle, we obtained the first four spectra
with airmass $<$1.1. The last spectrum taken on 1999 November 28, on the other hand,
might suffer from atmospheric refraction since we obtained it with
an airmass between 1.4-2.0. We took two 1200 sec exposures of the SNe per
grism, followed by a He-Ne arc lamp exposure, and spectra of
a flux standard with a 5 arcsec slit.

We obtained a spectrum of SN~1999ee on 1999 October 9 (two days after discovery) with the
R-C spectrograph of the CTIO~4-m telescope,
a 3K$\times$1K LORAL CCD, and grating KPGL-2 (316 lines mm$^{-1}$)
in first order. We took one 1800 sec exposure of the SN (through thick cirrus) with
a 1.5 arcsec slit oriented along the parallactic angle,
a He-Ar lamp exposure, and spectra of three flux standards through a 10 arcsec slit.
The resulting SN spectrum had a dispersion of 1.9~\AA~pix$^{-1}$ and
useful wavelength coverage of 0.33-0.87 $\mu m$. Second-order
contamination was expected beyond 0.66 $\mu m$ since we did not include
a blocking filter in the optical path. We obtained another spectrum
at CTIO with the 1.5-m telescope and the R-C spectrograph on 1999 October 27.
In this case we used a 1200$\times$800 LORAL CCD, grating 58 (400 lines mm$^{-1}$)
in second order, a 2 arcsec slit, and a CuSO$_4$ order-blocking filter.
The resulting spectrum had a dispersion of 1.1~\AA~pix$^{-1}$ and
useful wavelength coverage of 0.37-0.50 $\mu m$. We obtained six 1200 sec exposures
of SN~1999ee at an airmass $<$ 1.1, a He-Ar lamp image, and two spectra of flux standards. 

We also obtained a spectrum of SN~1999ee on 1999 October 16 with the
Las Campanas Dupont 2.5-m telescope and the Wide Field CCD Spectrograph.
We used a 2048$\times$2048 TEK CCD and a blue grism.
The resulting spectrum had a dispersion of 3~\AA~pix$^{-1}$ and
useful wavelength coverage of 0.36-0.92 $\mu m$. Second-order
contamination was expected beyond 0.66 $\mu m$ since we did not use
a blocking filter. 

The reductions consisted in subtracting the overscan and bias
from every frame. Next, we constructed a normalized flat-field
from the quartz-lamp image, duly normalized along
the dispersion axis. We proceeded by flat-fielding all of
the object frames and  extracting 1-D spectra from the
2-D images. We followed the same procedure for the He-Ar
frames which we used to derive the wavelength calibration
for the SNe. Then we derived a
response curve from the two flux stars, which we applied
to the SN spectra, in order to get flux calibrated spectra.
From the pair of flux-calibrated spectra that we obtained for each
spectral setup we removed cosmic rays and
obtained a clean spectrum of each SN. The last step consisted
in merging the spectra obtained with the different spectroscopic setups.
To avoid discontinuities  in the combined spectrum we grey-shifted the three spectra
relative to each other using the overlap regions.
Finally, we computed the synthetic $V$-band
magnitude from the resulting spectra  (following the precepts described
in Appendix B of \citet{hamuy01}) and we grey-shifted them so that the
flux level matched our observed $V$ magnitudes. We checked the
spectrophotometric quality of the spectra by computing 
$BRI$ synthetic magnitudes and comparing them to the
observed $BRI$ magnitudes of \cite{stritzinger02}. Excluding the spectra obtained on 1999 November 14,
this test yielded the following mean differences: $B_{obs}$-$B_{syn}$=-0.05$\pm$0.05,
$R_{obs}$-$R_{syn}$=0.00$\pm$0.07, and $I_{obs}$-$I_{syn}$=0.00$\pm$0.09,
which implies that the relative spectrophotometry at these wavelengths is accurate
to 10\% or better. The November 14 spectrum, on the other hand, shows $B$ fluxes 
that fall 0.35 mag lower than the photometric $B$ magnitudes and $I$ fluxes
that exceed by 0.2 mag the observed broadband $I$ magnitudes, perhaps
owing to atmospheric refraction effects.

\subsection{Infrared Spectroscopy}

We obtained five IR spectra with the VLT/Antu telescope at Cerro Paranal,
between 1999 October 9 and November 28 (see Table 1). We employed the IR spectro-imager ISAAC \citep{moorwood97}
in low resolution mode (R$\sim$500), with four different gratings that
permitted us to obtain spectra in the $SZ$, $J$, $H$, and $K$ bands. We used these gratings
in 5$^{th}$, 4$^{th}$, 3$^{rd}$ and 2$^{nd}$
order, respectively, which yielded useful data in the spectral ranges
0.984-1.136, 1.109-1.355, 1.415-1.818, and 1.846-2.560 $\mu m$.
The light detector was a Hawaii-Rockwell 1024x1024 array. 

A typical IR observation started during daytime by taking calibrations.
We began   taking flat-field images
using an internal source of continuum light. 
We obtained multiple on and off image pairs with
the same slit used during the night (0.6 arcsec).
We then took Xe-Ar lamp images (with the lamp on and off) with 
a narrow slit (0.3 arcsec) in order to map geometric distortions. 
The observations of the SNe
consisted in an ABBAAB cycle, where A is an image of the SNe
offset by 70 arcsec along the slit relative to the B image.
This technique of nodding the objects along the slit allowed
us to use the A image as an on-source observation and
the B image as the off-source sky frame, and viceversa.
At each position we exposed for 240 sec,
conveniently split into two 120-sec images in order to
remove cosmic rays and bad pixels from the final spectra.
After completing the ABBAAB cycle we immediately obtained
a pair of on-off arc lamp exposures 
without moving the telescope or changing optical elements to
ensure an accurate wavelength calibration. We then switched
to the next grating and repeated the above object-arc procedure
until completing the observations with the four setups.
For flux calibration we decided to observe a bright solar-analog star,
close in the sky to the SNe in order to minimize variations
in the atmospheric absorptions \citep{maiolino96}. The selected star was
Hip 109508, of spectral type G3V, $V$=8.0, $B$-$V$=0.59,
and located only 3$^\circ$ from the SNe. In this case we
took two AB pairs for each grating. To avoid saturating
the detector, we took the shortest possible exposures (1.77 sec)
allowed by the electronics that controlled the detector.
Since the minimum time required before offsetting the telescope
was $\sim$60 sec, we took ten exposures at each
position which provided an exceedingly good signal-to-noise ratio (S/N)
for the flux standard.

Our data reduction procedures were explained in detail by \citet{hamuy01},
so we include here only a brief summary. After dividing all of the object images
by a normalized flat-field, we performed a first-order sky subtraction
by subtracting the A images from the B exposures (and viceversa).
We then shifted the A-B image relative to the B-A image
in the spatial direction until matching the two spectra,
and we added the shifted A-B image to the B-A image so that
the resulting frame lacked any sky background, except for the
pixel-to-pixel fluctuations expected from photon statistics and
readout noise. We then extracted 1-D spectra of the objects from the
sky-subtracted frames, making sure to subtract residual DC offset and galaxy light from 
a window adjacent to the object. 
For flux calibration we adopted the technique described by \citet{maiolino96}, which consists
in dividing the spectrum of interest by a solar-type star to remove the strong
telluric IR features, and multiplying the resulting spectrum by the
solar spectrum to eliminate the intrinsic features (pseudonoise)
introduced by the solar-type star.
Before using the solar spectrum we convolved it with a
kernel function in order to reproduce the spectral resolution of the solar-analog
standard Hip 109508, we shifted it in wavelength
according to the radial velocity (68 km~s$^{-1}$) of Hip 109508, and
we scaled it down to the equivalent of
$V$=8.0 which corresponds to the observed magnitude of Hip 109508.
This technique worked very well to remove telluric lines. On the other hand,
it introduced a small systematic error in the flux calibration of the SN due
to departures between the solar spectrum and the actual spectral energy distribution
of the solar-analog standard. According to atmosphere models the difference
in continuum flux for stars with $T_{eff}$=5,500 and 6,000 K (which correspond to
spectral types G8V-F9V, \citet{gray94}) is smaller than 10\%
in the NIR region. A G3V star like Hip 109508 has nearly the same spectral type of
the Sun, so its effective temperature must be close (within $\pm$100 K)
to that of the adopted solar model. Hence, the flux difference between the
solar-analog standard and the adopted spectrum should be less than 10\%.
The $B-V$=0.59 color of Hip 109508 suggests little or no reddening so
the relative spectrophotometry is probably accurate to 5\% or better.

The result of these operations were four
spectra covering the $SZ$, $J$, $H$, and $K$ bands, which we
combined into one final IR spectrum for each SN. Given the significant overlap of
the $SZ$ and $J$ band spectra, we were able to grey-shift the $J$
spectrum relative to the $SZ$ spectrum. Then we used our broad-band $JHK$ magnitudes
to grey-shift the individual spectra.

On six occasions we used the NTT/SOFI instrument at La Silla in service observing
mode, in order to complement the IR spectroscopic follow-up. We employed 
a Hawaii-Rockwell 1024x1024 light detector, and the blue
and red grisms that permitted us to obtain spectra in the ranges 0.95-1.64 $\mu m$ and 1.53-2.52 $\mu m$,
with resolutions of 7.0 and 10.2 \AA, respectively. We obtained the spectra
with slit widths between 0.6 and 1 arcsec. We nodded the objects along 
the slit, after exposing for 150 sec at each position. We completed several pairs
of AB observations for the SNe, we took Xe-Ne arc images at the position of
the targets, and we observed the flux standard Hip 109508.
We reduced these data following the same procedure explained above for ISAAC.
The large overlap of 1.53-1.64 $\mu m$ between the blue and red spectra 
allowed us to grey-shift the red spectra relative to blue spectra, except on
1999 October 18 when we were only able to obtain the red spectrum. Finally, we shifted the
resulting spectra using the observed $JHK$ magnitudes \citep{krisciunas02}.

\section{Results}

\subsection{SN~1999ee}

Before analyzing the spectroscopic observations it proves necessary to mention
the photometric properties of SN~1999ee. A detailed discussion of the photometric
observations can be found in \citet{stritzinger02}. In brief, the decline rate 
of $\Delta$$m_{15}(B)$=0.94 puts SN~1999ee in the group of slowest-declining SNe~Ia.
After correcting the observed magnitudes for $K$ terms and dust extinction, 
and assuming a Hubble constant of 63 km~s$^{-1}$~Mpc$^{-1}$ \citep{phillips99}, SN~1999ee
had a peak visual magnitude of $M_V$=-19.94 (JD 2451469.1).
These values lie comfortably well with the peak magnitude-decline rate
relation for SNe~Ia \citep{phillips99}. This analysis reveals that SN~1999ee
was a luminous Type Ia event and a very interesting object to check the claim
by \citet{li01b} that slow-declining events display spectroscopic
peculiarities before maximum light.

\begin{figure}
\figurenum{1}
\epsscale{1.0}
\plotone{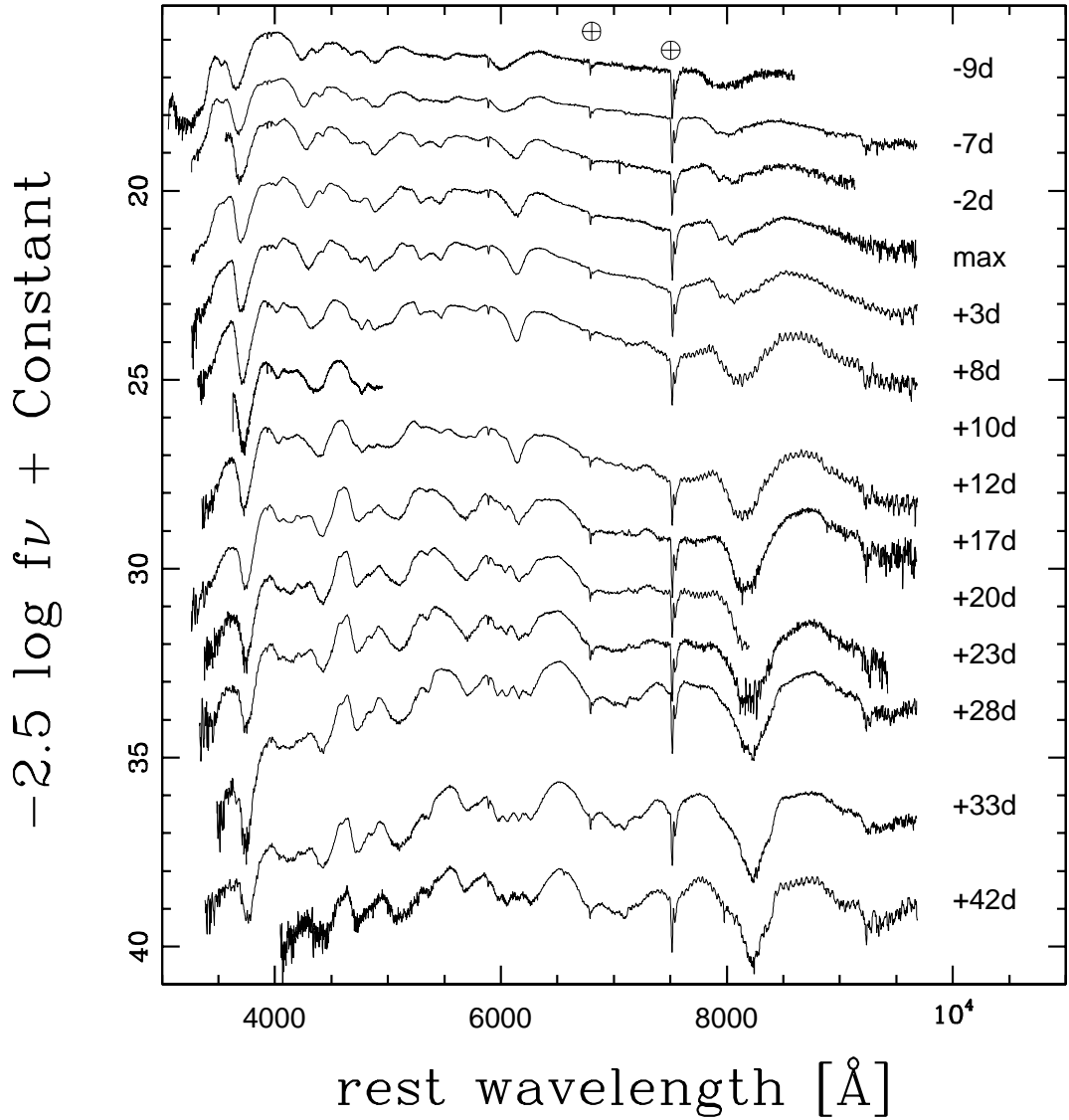}
\caption{Optical spectroscopic evolution of SN~1999ee
in AB magnitudes. Time (in days) since $B$ maximum is indicated for each spectrum.
The $\oplus$ symbols show the main telluric features. 
\label{sn99ee.opt.fig}}
\end{figure}

Figure \ref{sn99ee.opt.fig} displays the optical spectra of SN~1999ee in the
rest-frame of the SN, after correcting the observed spectra for the 3,498 km~s$^{-1}$ recession velocity
of the host galaxy. We also dereddened the spectra assuming a Galactic reddening
of $E(B-V)$=0.02 \citep{schlegel98}, and a host galaxy reddening of $E(B-V)$=0.30 derived
from the SN optical colors \citep{stritzinger02} and optical/IR colors \citep{krisciunas02}.
The strongest telluric lines are indicated with the $\oplus$ symbol.
The first spectrum, taken on JD 2451460.56 (9 days before maximum), exhibited a blue continuum
and the characteristic P-Cygni profile of the Si II $\lambda$6355 line of SNe~Ia.
Other prominent features in this spectrum were the absorption at $\sim$3200 \AA~
attributed to Co II, the Ca II H\&K $\lambda\lambda$3934,3968 blend,
Mg II $\lambda$4481, the blend of lines attributed to Fe II, Si II, S II around 4550-5050 \AA~, and the
Ca II $\lambda\lambda$8498,8542,8662 triplet.
The presence of the Na I D $\lambda\lambda$5890,5896, Ca II $\lambda$3934, and
Ca II $\lambda$3968 interstellar lines with equivalent widths of 2.3, 0.8, and 0.6~\AA, respectively,
revealed a non-negligible amount of absorption in the host galaxy, which agrees
with the color analysis of SN~1999ee.
By maximum light the ``W'' shaped S II $\lambda\lambda$5454,5640 feature was well developed
and the Ca lines were significantly stronger. Si II $\lambda$4129 was already present
although it was quite weak. The Si II $\lambda$5972 line also was weak relative to Si II $\lambda$6355.
Nugent et al. (1995) defined the parameter ${\cal R}$(Si~II) as the
relative fluxes of this pair of lines and found a tight correlation with the SN luminosity.
In this case we obtained ${\cal R}$(Si~II)=0.23 which corresponds to the value of a
luminous SNe~Ia, in good concordance with the decline rate of $\Delta$$m_{15}(B)$=0.94
and the peak magnitudes derived from the photometric observations.
By day 20 the spectrum was much redder. The Si II $\lambda$6355 line
was very weak, the Ca triplet had a pronounced P-Cygni profile, and
the region between 4000-5500 \AA~was dominated by strong
features due to Fe II transitions. The absorption attributed
to Na I D $\lambda\lambda$5890,5896 was very prominent. 
Overall, these data showed that SN~1999ee was a genuine Type Ia event
with the usual spectral features of other normal SNe~Ia, usually
known as Branch-normal \citep{branch93}.

\begin{figure}
\figurenum{2}
\epsscale{1.0}
\plotone{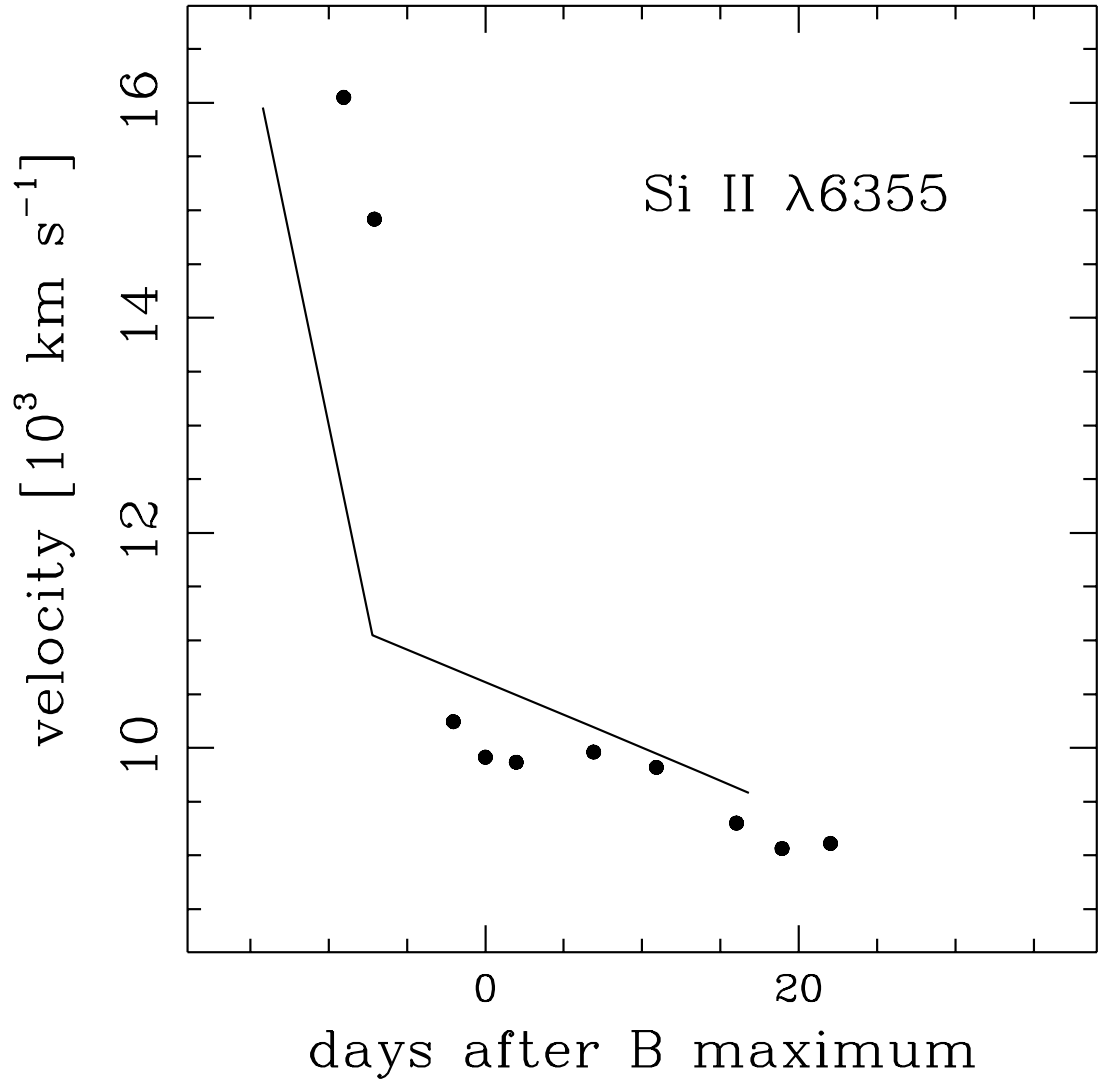}
\caption{Expansion velocities derived from the absorption
minima of Si II $\lambda$6355 for SN~1999ee, after correcting for the recession
velocity of the host galaxy. The solid line corresponds to SN~1990N. 
\label{sn99ee.vel.fig}}
\end{figure}

Figure \ref{sn99ee.vel.fig} presents the expansion velocities derived from the absorption
minima of Si II $\lambda$6355, after correcting for the recession
velocity of the host galaxy. Apparently the velocity derived from Si II $\lambda$6355
decreased rapidly from 16,000 to 10,000 km s$^{-1}$ between days -7 and -2,
after which the velocity decreased slowly.
For comparison we include as solid line 
the velocities for SN~1990N which was also caught several days before
maximum \citep{leibundgut91}. Both SNe showed an  inflection in
the velocity curve. Before maximum, however, the expansion
of SN~1999ee was somewhat higher compared to contemporary velocities 
of SN~1990N. At maximum light and later times, on the other hand,
both SNe had similar velocities.

\begin{figure}
\figurenum{3}
\epsscale{1.0}
\plotone{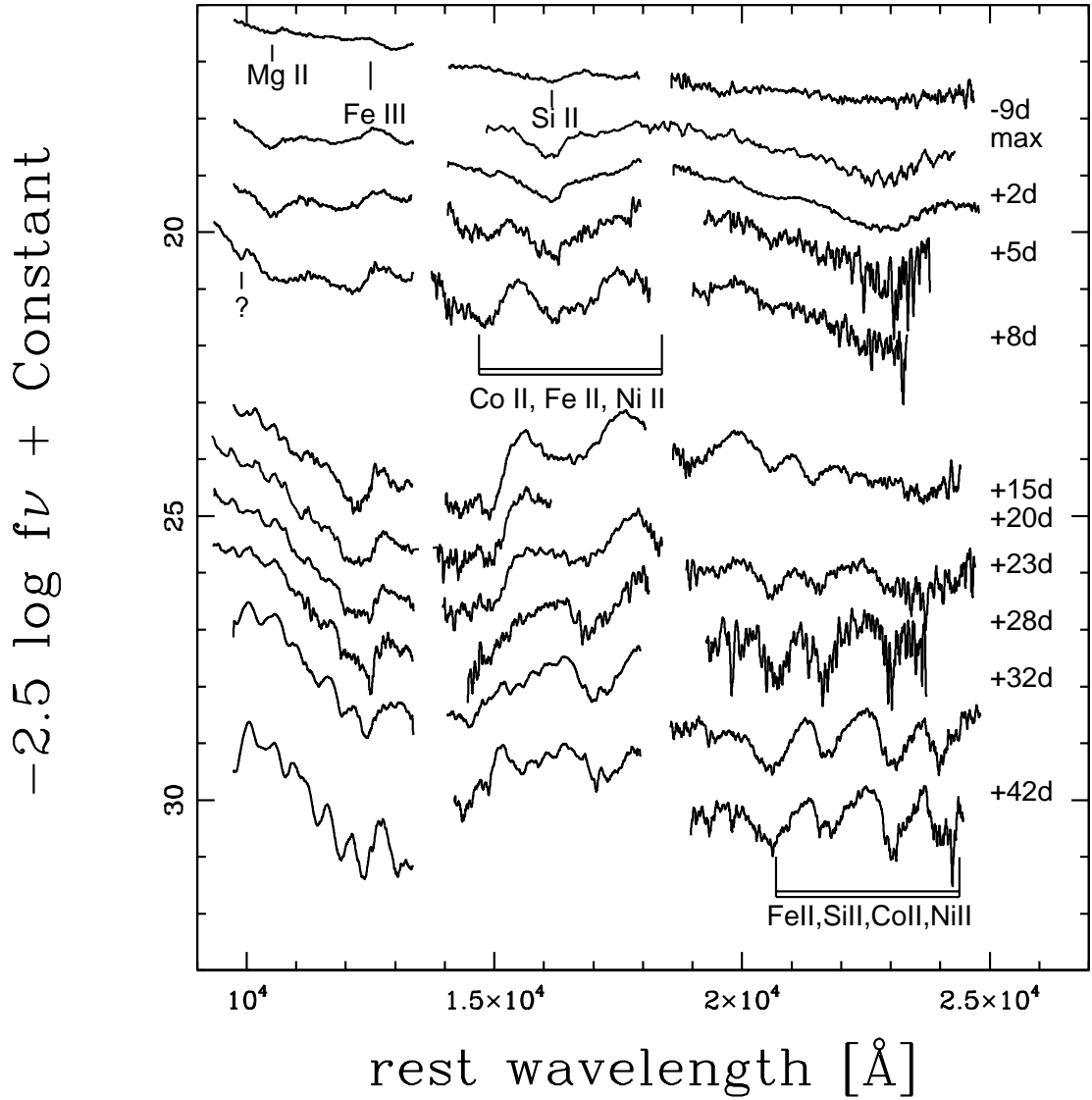}   
\caption{IR spectroscopic evolution of SN~1999ee
in AB magnitudes. The most prominent features are labeled.
Time (in days) since $B$ maximum is indicated for each spectrum.
\label{sn99ee.ir1.fig}}
\end{figure}

Figure \ref{sn99ee.ir1.fig} shows the resulting rest-frame IR spectra of SN~1999ee. 
The first spectrum, taken nine days before maximum, was quite featureless.
A weak absorption line could be seen at 10520 \AA~with a weak
emission component that seemed to constitute a P-Cygni profile. This feature
was observed for the first time in an early-time spectrum of SN~1994D and was
tentatively identified as due to either He I $\lambda$10830 or Mg II $\lambda$10926 \citep{meikle96}.
According to the atmosphere models computed by \citet{wheeler98}, this
feature should be due to Mg II since the amount of He in their models is
not sufficient to form a line. This implies that the minimum of
the Mg absorption had an expansion velocity of 11,100 km~s$^{-1}$. 
An emission due to Fe III \citep{rudy02} was also present around 12550 \AA.
There was an indication of a P-Cygni profile with the absorption minimum 
at 16160 \AA~ and peak emission at 16840 \AA~. This feature  was
also present in the SN~1994D spectrum.
The theoretical modeling of Wheeler et al. suggests that 
this line is  due to Si II, although it might have a Fe II component \citep{axelrod80}.
At longer wavelengths the spectrum is even more featureless except
for a small P-Cygni line with a minimum at 19590 \AA~ and
peak at 20440 \AA~. The emission component was also noted in
SN~1994D and was attributed to Si III by Wheeler et al.

By maximum light the Mg II $\lambda$10926, Fe III, and Si II feature
near 16500 \AA~became more prominent and a peak developed
around 18100 \AA. One week past maximum the spectrum 
was dramatically different and became dominated by strong absorption/emission
features.  The Mg II $\lambda$10926 line disappeared 
and a weak feature at 10000 \AA~ became evident, possibly
forming a P-Cygni profile. The most remarkable features were the strong and
wide (FWHM $\sim$ 12,000 km~s$^{-1}$) peaks
attributed to blends of Co II, Fe II, and Ni II at 15500 and 17500 \AA~\citep{wheeler98}.
Two weeks after maximum these peaks developed even further and new peaks
appeared  at 19900 and 21000 \AA.
A series of unidentified absorptions also could be observed in the $J$ band at 10400, 10700, and 11100 \AA.
One month after maximum two additional broad emission peaks became
evident at 22500 and 23600 \AA~which, according to Wheeler et al.,
are caused by iron-group elements (Co, Ni) as well as
intermediate mass elements with appreciable contribution from Si.
While the broad emissions appeared to redshift significantly as the SN evolved,
the features that we identify as absorptions did not show any appreciable
change in velocity.

\subsection{SN~1999ex}

\begin{figure}
\figurenum{4}
\epsscale{1.0}
\plotone{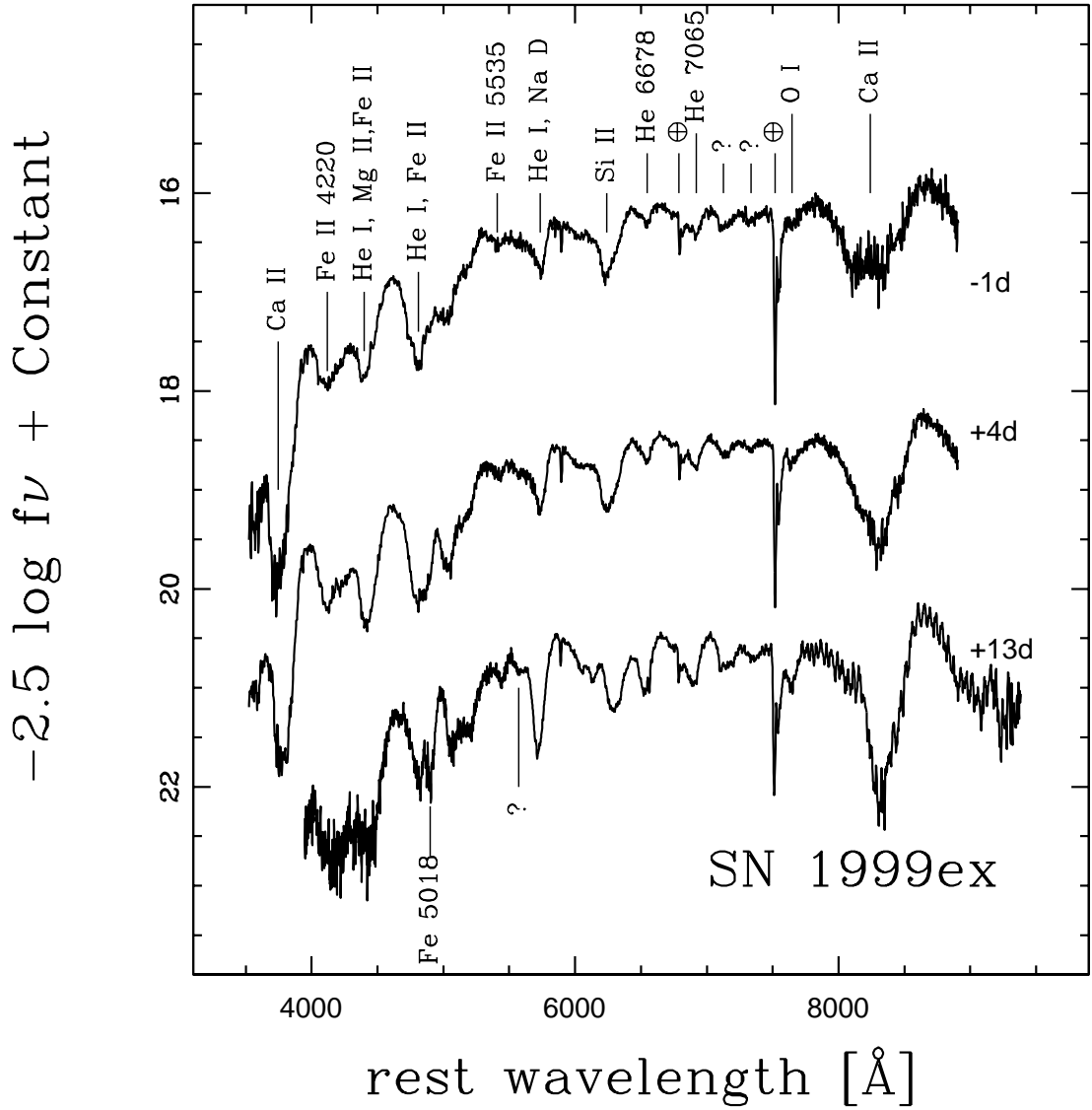}
\figcaption{Optical spectroscopic evolution of SN~1999ex
in AB magnitudes.  Time (in days) since $B$ maximum is indicated for each spectrum.
The most prominent features are labeled as well as the main telluric features with
the $\oplus$ symbol.
\label{sn99ex.opt.fig}}
\end{figure}

Figure \ref{sn99ex.opt.fig} displays the optical spectra of SN~1999ex in the
SN rest-frame. The first spectrum, obtained on JD 2451496.62 (one
day before $B$ maximum), was characterized by a reddish continuum and
several broad absorption/emission features due to Ca II H\&K $\lambda\lambda$3934,3968,
Fe II, Mg I, Na I D $\lambda\lambda$5890,5896, Si II $\lambda$6355, and the Ca triplet with a clear
P-Cygni profile. We also found convincing evidence for He I $\lambda\lambda$6678,7065
with expansion velocities of 6,000 km~s$^{-1}$ (similar to velocities displayed
by the Fe, Mg, Na, Si, and O lines).
It is possible that the feature at $\sim$4400 \AA~, usually  attributed to a blend
of Mg II and Fe II, had a contribution due to the He I $\lambda$4471. Likewise,
the feature at $\sim$4800 \AA~could be a blend of Fe II $\lambda$4924 and He I $\lambda$4921,
while the Na I D $\lambda\lambda$5890,5896 doublet could have a contribution from He I $\lambda$5876.
Interstellar lines due to Na I D $\lambda\lambda$5890,5896, Ca II $\lambda$3934, and
Ca II $\lambda$3968 were observed in absorption with equivalent widths of 2.8, 1.8, 0.9~\AA, 
at the wavelengths corresponding to the rest-frame of the host galaxy.
According to the correlation between equivalent widths and reddening for Galactic stars
\citep{munari97}, SN~1999ex was reddened by $E(B-V)$$\sim$1.0$\pm0.15$. However,
there appears to be a significant departure of SNe from this calibration \citep{hamuy01},
so this estimate is very uncertain.

The first spectrum of SN~1999ex bore quite resemblance to that of SN~1994I.
\cite{filippenko95} argued that SN~1994I was a Type Ic event
on the grounds that this object
1) did not show evidence for hydrogen lines (ruling out a Type II event), 
2) had a Si II $\lambda$6355 line more subtle than in classical SNe Ia, and
3) He I $\lambda$5876 was absent or weak (and blended with Na D) (ruling out a Type Ib event).
Based on the resemblance between these two spectra, we initially classified
SN~1999ex as a Type Ic SN \citep{hamuy99}. However, the relatively stronger He lines
displayed by SN~1999ex suggest that this object could well be an intermediate case between
the Ib and Ic subclasses (we will return to this point in section \ref{sn99ex.sec}).
During the following two weeks the SN spectrum evolved slowly and 
most of the lines became stronger.

\begin{figure}
\figurenum{5}
\epsscale{1.0}
\plotone{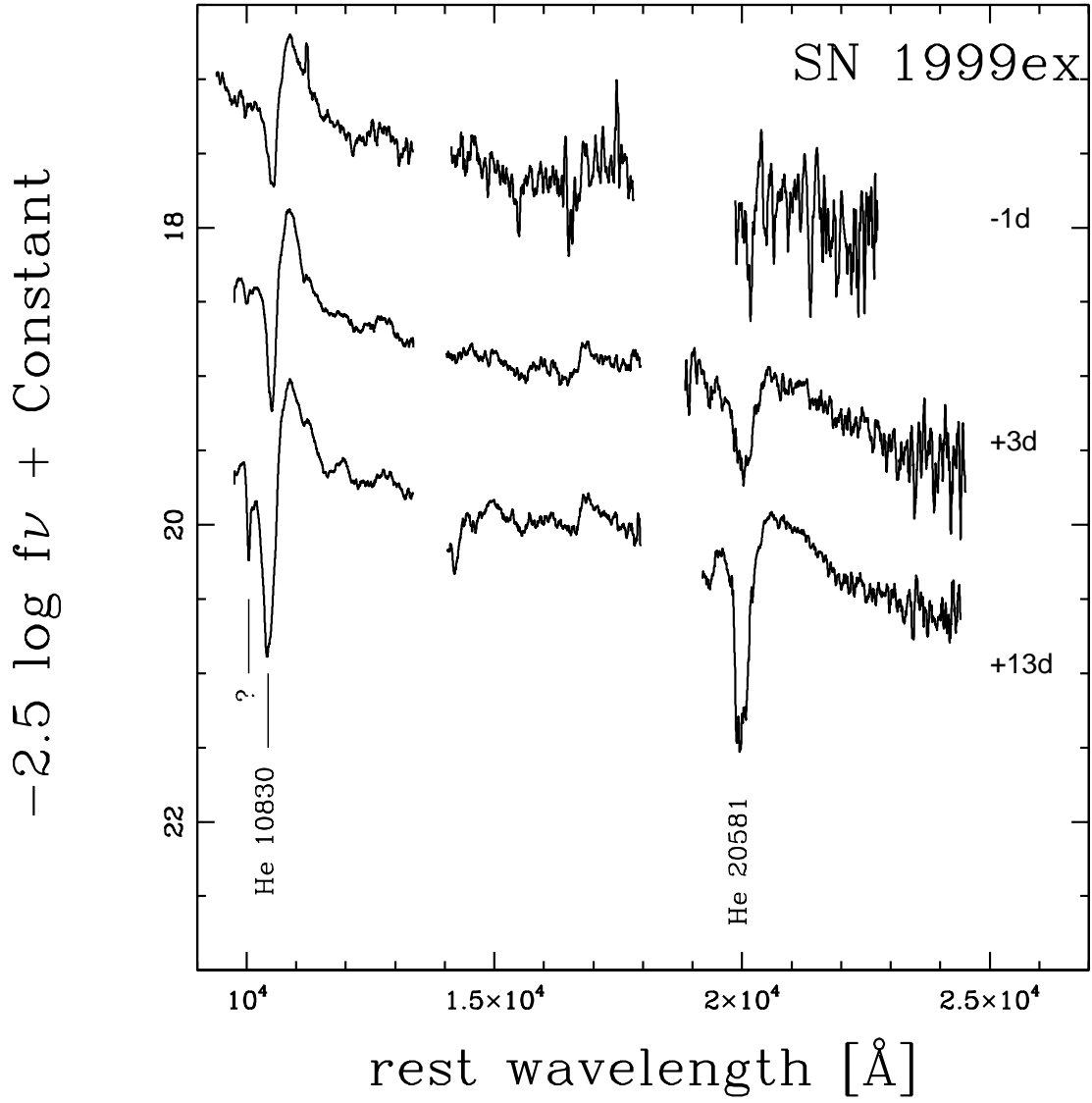}
\figcaption{IR spectroscopic evolution of SN~1999ex
in AB magnitudes. The two most prominent features are due to He I, while
the absorption at 10,000 \AA~remains unidentified. 
Time (in days) since $B$ maximum is indicated for each spectrum.
\label{sn99ex.ir.fig}}
\end{figure}

Figure \ref{sn99ex.ir.fig} displays the IR spectra of SN~1999ex in the
SN rest-frame. The most prominent feature in these spectra
is the P-Cygni profile of He I $\lambda$10830 which was also observed
in SN~1994I.  The minimum of the absorption
yields an expansion velocity of 8,000 km~s$^{-1}$, in good
agreement with the values derived from the optical spectra. The other evident feature
is the  P-Cygni profile due to He I $\lambda$20581, with the same
expansion velocity. An unidentified absorption at 9970-10045 \AA~
can be clearly seen, which should have a rest wavelength between 10200-10260 \AA.
A few weak unidentified lines can be seen also in the $H$ band.

\section{Discussion}

\subsection{SN~1999ee}

Although Type Ia SNe display a large degree of spectroscopic homogeneity,
there are several examples of SNe with spectral peculiarities.
Different subclasses have been identified including
SN~1991T-like objects, Branch-normal SNe, and SN~1991bg-like events \citep{branch93}.
Recently, the Lick Observatory Supernova Survey (LOSS) produced
a SN sample with well-understood selection effects \citep{li01a} which permitted
them to assess the intrinsic peculiarity rate of SNe~Ia \citep{li01b}. This study showed
that the sample of nearby SNe~Ia comprises 20\%, 64\%, and 16\% of 1991T-like, normal,
and 1991bg-like objects, respectively.

\citet{li01a} defined a normal SN as one with prominent features due to Si II $\lambda$6355, 
Ca II H\&K $\lambda\lambda$3934,3968, as well as additional lines of S II, O I, and Mg II
around maximum light.
The designation SN~1991T in such work is not strictly correct since the prototype of the
1991T class defined by \citet{li01b} is SN~1999aa. While SN~1999aa's main difference relative
to normal SNe is just a relative weakness of the Si II $\lambda$6355 feature before maximum
that becomes almost indistinguishable from normal SNe after peak (see Fig. 5 of \citet{li01b}),
SN~1991T was a much more extreme event in the sense that the pre-maximum spectrum did not display
the Ca II H\&K $\lambda\lambda$3934,3968 blend shown by Branch-normal SNe, and also because
it remained spectroscopically distinct several days after maximum \citep{phillips92}.
For the sake of clarity, in what follows we refer to the 1991T-like events of Li et al.
as 1999aa-like objects. The distinguishing feature of SN~1991bg-like objects is
the Ti II absorption around 4100-4000 \AA. There are many examples of intermediate
class objects, which suggests a continuous spectral sequence among SNe~Ia.

\begin{figure}
\figurenum{6}
\epsscale{1.0}
\plotone{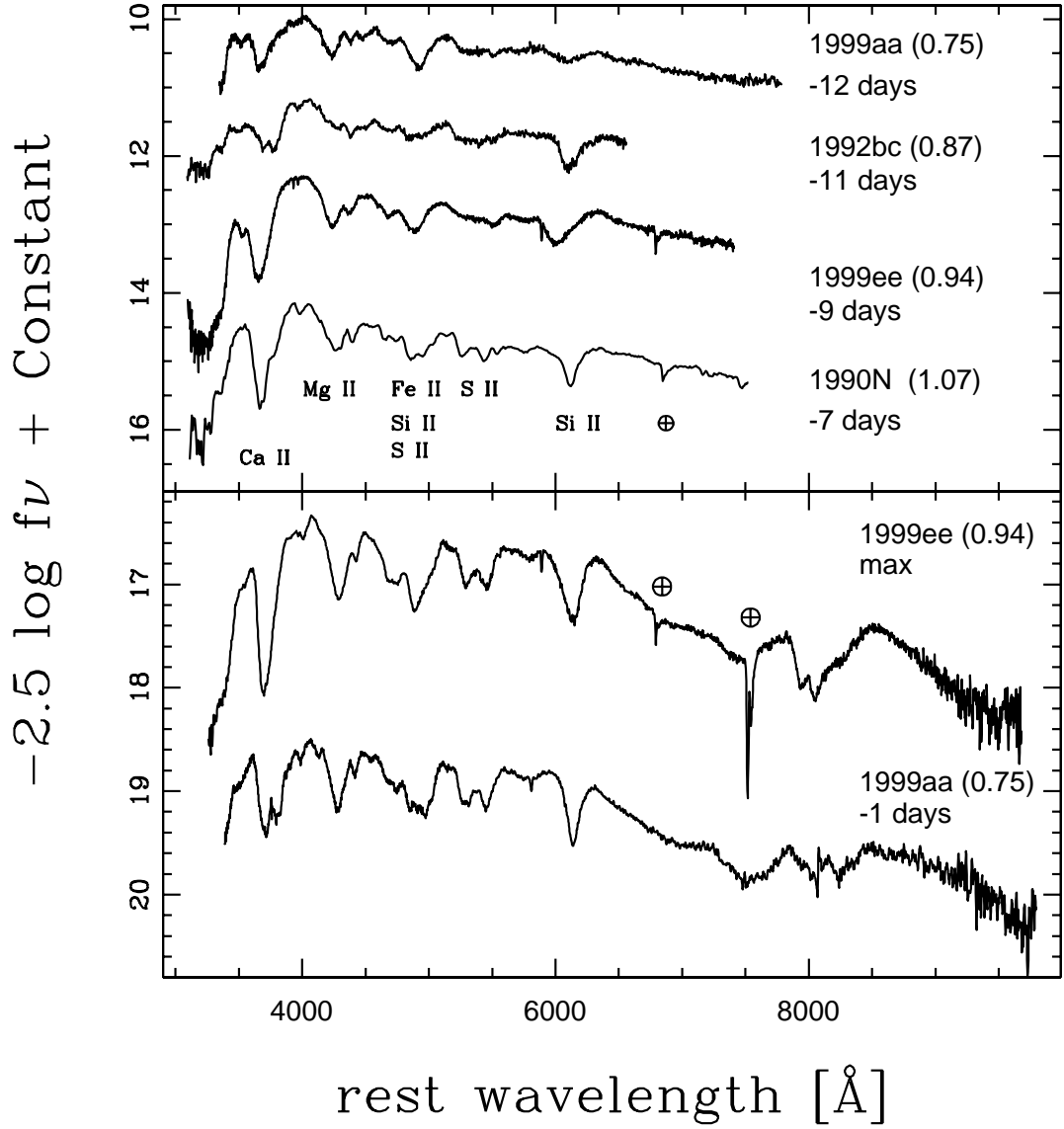}     
\figcaption{(top) Comparison of the optical spectra of three
slow-declining SNe~Ia and the Branch-normal SN~1990N obtained about 10 days before $B$ maximum (top).
The spectra have been shifted with respect to each other by arbitrary amounts  to facilitate the comparison.
In parenthesis are given the decline rates $\Delta$$m_{15}(B)$ for each SN.
(bottom) Same as above but for SN~1999aa and SN~1999ee near maximum light.
\label{sn99ee.opt.comp.fig}}
\end{figure}

Because 1999aa-like events display slow-declining lightcurves ($\Delta$m$_{15}$(B)$\leq$1.0)
and high peak luminosities,
and 1991bg-like SNe show fast-decline rates ($\Delta$m$_{15}$(B)$\geq$1.7) and low peak luminosities,
it is interesting
to ask if the photometric behavior can be used to predict spectroscopic peculiarities.
As a slow-declining ($\Delta$m$_{15}$(B)=0.94) and luminous ($M_V$=-19.94) event observed
well before maximum light, SN~1999ee provides a good opportunity to investigate whether
or not all luminous SNe display spectroscopic peculiarities before peak.
To examine this point in detail Figure \ref{sn99ee.opt.comp.fig} (top) shows
a comparison of the spectrum of SN~1999ee taken 9 days before $B$ maximum to that
of three other SNe including SN~1999aa with a decline rate of $\Delta$$m_{15}(B)$=0.75
\citep{krisciunas00}, the Cal\'an/Tololo SN~1992bc with $\Delta$$m_{15}(B)$=0.87 \citep{hamuy96},
and the normal SN~1990N with $\Delta$$m_{15}(B)$=1.07 \citep{leibundgut91,lira98}.
While SN~1999aa lacked the Si II $\lambda$6355 feature compared to the Branch-normal SN~1990N,
both SN~1992bc and SN~1999ee displayed strong and well-defined Si features
as early as ten days before peak.  This shows that
{\em not all slow-decliners are spectroscopically peculiar before maximum light}.
As mentioned above, 1999aa-like events show a normal spectrum at peak brightness, which
can be clearly appreciated in the bottom panel of Figure \ref{sn99ee.opt.comp.fig} from
the spectra of SN~1999aa and SN~1999ee. The fact that the spectral differences displayed
by the luminous SNe are only limited to the very first days of evolution suggests
that the 1999aa-like and the Branch-normal SNe (represented
here by SN 1999ee) are apparently similar explosions and that spectral
diversity at early times probably reflects small differences in mixing or
other chaotic behaviors rather than fundamental differences in the
character of the explosions.

There are a handful of IR spectra published for SNe~Ia \citep{meikle96,bowers97,hernandez00,jha99,rudy02}.
\cite{hoflich02} recently observed SN~1999by and assembled the first continuous sequence
of IR spectra for a SN~Ia, extending from 3 days before maximum through day +15 after peak.
To our knowledge the IR spectra of SN~1999ee -- extending from day -9 to +42 --
constitute the most complete IR spectroscopic sequence of a SN~Ia and
its Branch-normal nature makes it an ideal object in which to study the homogeneity
of this class of objects at these wavelengths. We proceed now to address this issue
from a sample of five SNe~Ia encompassing a wide range in decline rates and optical
spectroscopic properties. The sample comprises the Branch-normal SN~1999ee, SN~1998bu \citep{jha99,hernandez00},
and SN~1994D \citep{meikle96}, with $\Delta$$m_{15}(B)$=0.94,1.01,1.32, respectively \citep{phillips99}.
\footnote{We include in this comparison two spectra of SN~1998bu obtained by one of us
(R.D.B.) with the CTIO 4-m Infrared  Spectrograph on 1998 May 16 (3 days before B maximum)
and June 3 (15 days after peak).}
We also include SN~1999by ($\Delta$$m_{15}(B)$=1.90) which had optical spectra
that showed resemblance to SN~1991bg \citep{garnavich01}, and SN~2000cx
($\Delta$$m_{15}(B)$=0.93) which shared some of the spectroscopic
peculiarities displayed by SN~1999aa-like events (e.g. weak Si II $\lambda$6355)
before maximum and a ``sui generis'' post-maximum behavior owing
to unusually strong Fe III and weak Fe II lines \citep{li01c}.

\begin{figure}
\figurenum{7}
\epsscale{1.0}
\plotone{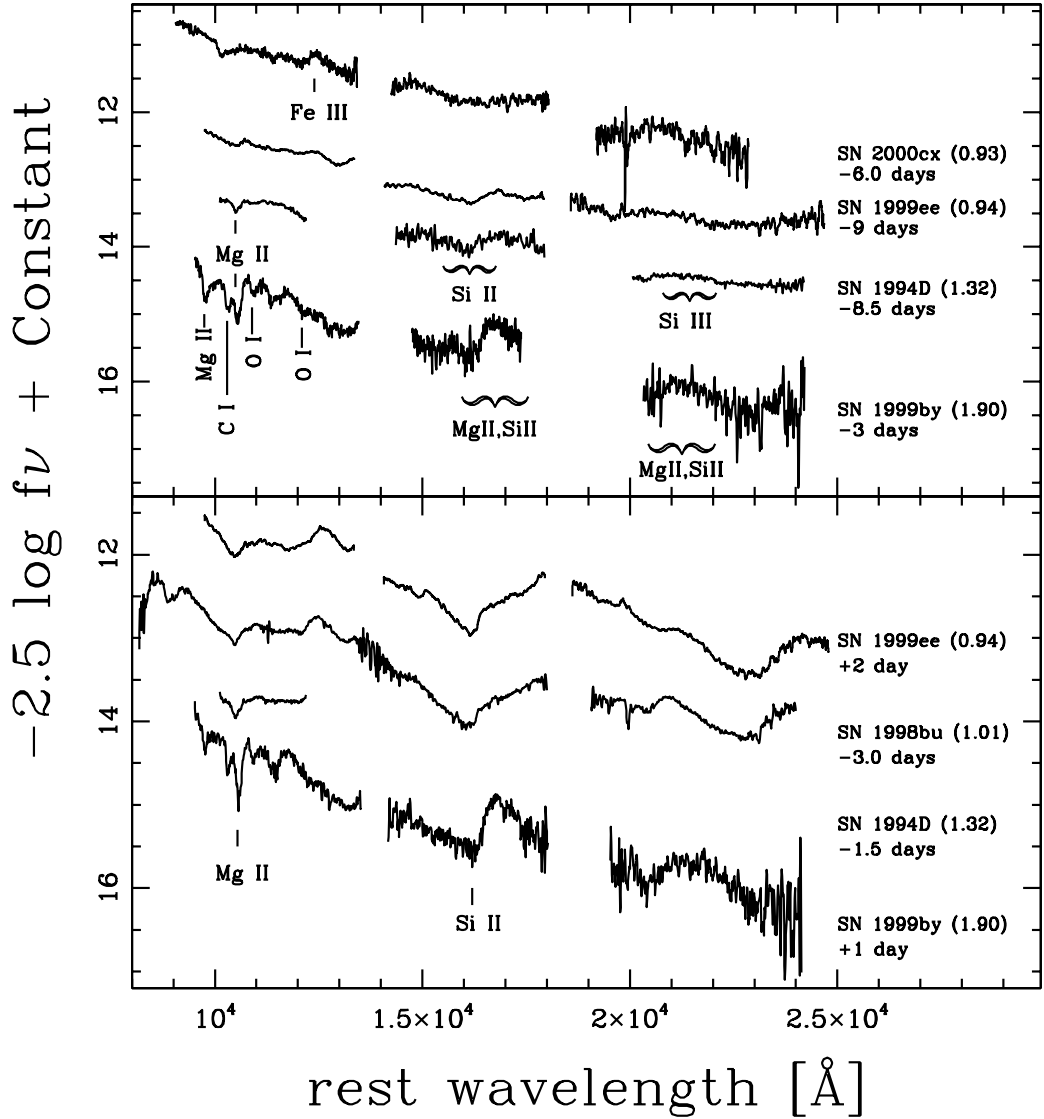}     
\figcaption{Comparison of the pre-maximum and maximum-light
IR spectra of SNe~1994D, 1998bu, 1999by, 1999ee, and 2000cx.
Time (in days) since $B$ maximum is indicated for each spectrum along
with the decline rates $\Delta$$m_{15}(B)$.
\label{sn99ee.ir.comp3.fig}}
\end{figure}

Figure \ref{sn99ee.ir.comp3.fig} (top) compares pre-maximum 
spectra of SN~2000cx, 1999ee, 1994D, and 1999by. The resemblance between
the two Branch-normal SNe (1999ee and 1994D) is remarkable. The largest, yet subtle, difference
was the Mg II $\lambda$10926 feature which was narrower and deeper in SN~1994D. The expansion
velocities derived from the absorption minimum was quite similar, $\sim$10,500 km~s$^{-1}$.
Both the Si II and Si III features near 16500 \AA~ and 20000 \AA~were clearly present on both SNe.
With the exception of the high Mg II velocity ($\sim$20,000 km~s$^{-1}$), SN~2000cx showed 
a normal featureless spectrum.
Since magnesium is destroyed by oxygen burning, it is expected in
material which experiences burning at the lowest densities near the
surface. The high Mg velocity led \cite{rudy02} to conclude that
nuclear burning in SN~2000cx extended farther out than in normal SNe,
though it is equally likely that in this supernovae a blob containing
carbon burning products drifted closer to the surface than in some
other SNe~Ia. The greater luminosity and slower decline of SN~2000cx
argue for greater $^{56}$Ni production, and this would manifest itself also
in greater excitation in the ejecta. Indeed, SN~2000cx showed an
emission feature at 12500 \AA~attributed to Fe III $\lambda\lambda$12786,12920,13003 \citep{rudy02},
which was also present in the normal SN~1999ee but at a lower strength, a weaker than
normal Si II feature near 16500 \AA, and a stronger Si III line around 20000 \AA.

SN~1999by showed a pre-maximum spectrum noticeably different than that of normal SNe.
The main difference was due to C I and O I lines, which were clearly absent in the
spectra of other SNe. \citet{hoflich02} suggested that nuclear burning in SN~1999by did not
extend as far out as in normal SNe~Ia. It is possible, however, that
the presence of the C I and O I lines was due to a lower excitation evidenced by
the smaller $^{56}$Ni production, the low luminosity and fast decline of SN~1999by.
The spectral differences are particularly evident
in the $J$ band. The Mg II $\lambda$10926 feature was deeper than normal, although its
velocity of $\sim$10,000 km~s$^{-1}$ was similar to that observed in normal SNe.
In the $H$ band SN~1999by showed a strong broad emission feature
due to a blend of Si II and Mg II lines which was also present in the other SNe but with a
lower strength, presumably due to a weaker contribution from Mg II. The $K$ band spectrum
is also bumpier than normal due to Mg II lines.

Around maximum light the Mg II $\lambda$10926 feature was present in all three
normal SNe (1999ee, 1998bu, and 1994D) with similar strengths and profiles
(bottom panel of Figure \ref{sn99ee.ir.comp3.fig}).
\cite{meikle96} noted that the absorption minimum of the Mg II line did not show a shift
in wavelength during the pre-maximum evolution of SN~1994D, as opposed to the optical
lines which all redshifted with time as the photosphere receded through
the ejecta.  We confirmed the absence of any significant Doppler shift in Mg II
from SN~1999ee which, according to \citet{wheeler98} was due to the fact that
the photosphere had already receded below the inner edge of the magnesium layer
at this early phase. 
It is interesting to note the absence in these spectra
of the Ca II line near 11500 \AA, a feature predicted by the models of \citet{wheeler98}
that proves to be a useful diagnostic of the location of the transition layer
between complete and incomplete silicon burning.
The Si II absorption near 16500 \AA~and the spectral features in the $K$ band were visible
in the normal SNe~1999ee and 1998bu with a high degree of similarity.
The pre-maximum spectral peculiarities of SN~1999by mentioned above
were still evident near maximum light.

\begin{figure}
\figurenum{8}
\epsscale{1.0}
\plotone{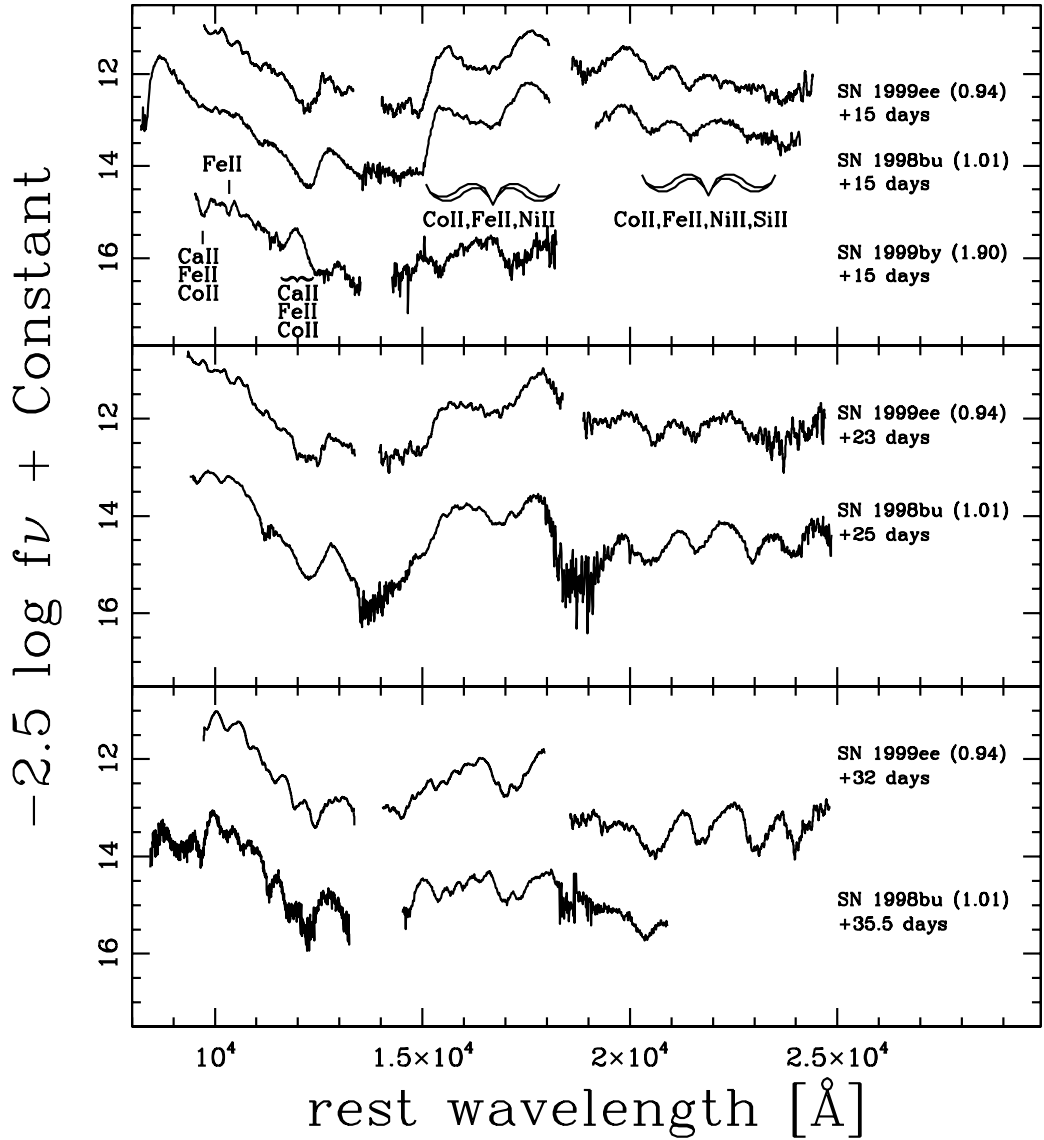}     
\figcaption{Comparison of the post-maximum IR spectra of SNe~1998bu, 1999by, and 1999ee.
Time (in days) since $B$ maximum is indicated for each spectrum along
with the decline rates $\Delta$$m_{15}(B)$.
\label{sn99ee.ir.comp1.fig}}
\end{figure}

Figure \ref{sn99ee.ir.comp1.fig} compares the post-maximum spectra available to us.
Two weeks after peak the two normal SNe~1999ee and 1998bu displayed an impressive
spectral homogeneity with strong peaks and valleys at 15000-17000 \AA~and
22000-26000 \AA~due to Fe II, Co II, Ni II, and Si II. The presence of
Fe II, Co II, Ni II, and Si II is due to the ionization dropping predominantly to
these ions, which happen to have strong lines in the near IR.
The spectral differences of the fast-decliner SN~1999by persisted at
this epoch. While its $J$ spectrum had contributions from Ca II lines \citep{hoflich02}
that were absent in SNe~1999ee and 1998bu, the broad emission around 15000-17000 \AA~was
weaker than normal. At later epochs, the spectra of SNe~1999ee and 1998bu 
became much more complicated, yet their similarity remained remarkable.

\begin{figure}
\figurenum{9}
\epsscale{1.0}
\plotone{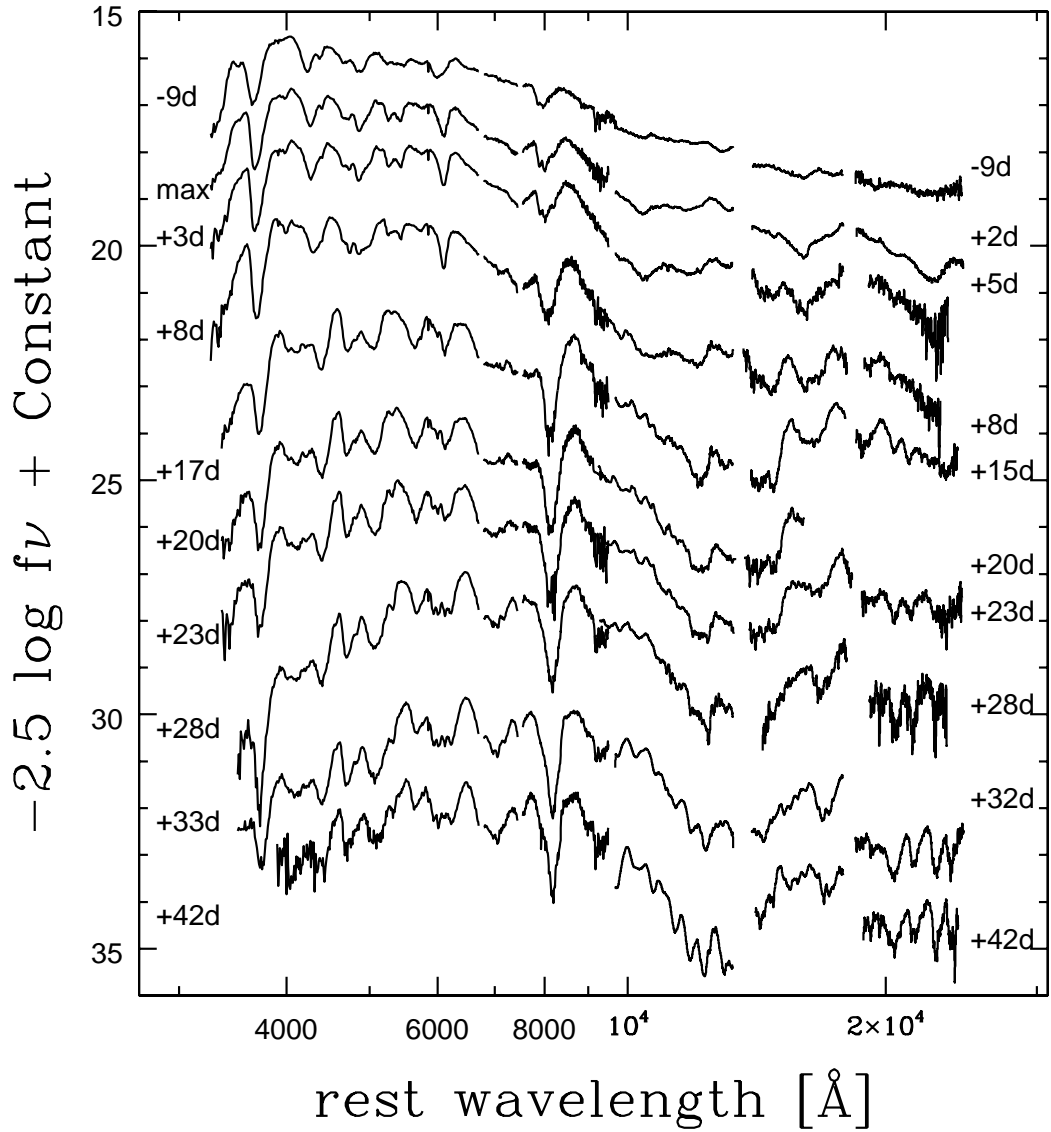}
\figcaption{Combined optical and IR spectra of SN~1999ee in AB magnitudes. 
Time (in days) since $B$ maximum is indicated for each spectrum.
\label{sn99ee.fig}}
\end{figure}

As a result of this observational campaign we obtained the most
complete optical/IR observations of a Type Ia SN, with an unprecedented
wavelength and temporal coverage beginning nine days before maximum light.
Moreover, since we were able to obtain the IR spectra within one or two days
from the optical spectra, it was possible to combine these observations,
as shown in Figure \ref{sn99ee.fig}. 
This exercise revealed the excellent agreement between the optical and IR fluxes,
a result that proved possible by synthesizing broad-band magnitudes
and adjusting the flux scales according to the observed magnitudes.
This optical/IR sequence shows that, while the pre-maximum optical spectrum
was dominated by strong P-Cygni profiles of intermediate mass elements
like Ca II, Si II, Mg II, S II, the IR was characterized by a smooth, almost featureless continuum.
By maximum light, on the other hand, the IR spectrum was already
dominated by broad features, and one week later the IR flux was
mostly powered by emission lines of iron group elements like Co  II, Fe II, and Ni II
(freshly synthesized in the innermost layers of the ejecta) that were particularly
prominent in the $H$ and $K$ bands. At the same time a dramatic flux dip around 12000 \AA~began
to develop.  Previous spectroscopic observations have already revealed this $J$-band
deficit in other SNe~Ia \citep{bowers97}, which is responsible for the red $J-H$
color displayed by SNe~Ia after maximum light \citep{elias85}.

This spectral sequence lends support to the suggestion by \citet{spyromilio94} and
\citet{pinto00} that the ``photosphere'' recedes rapidly to the center of the supernova in the IR
while at optical wavelengths the greater opacity arising from the
higher spectral density of lines keeps the photosphere at higher
velocities.  Emission at longer wavelengths increases after maximum
light due to a shift in ionization to more-neutral species which have
greater emissivity in the near IR. In this model, the $J$-band
deficit is due to the relative absence of lines (and hence opacity)
in the 12000 \AA~region rather than increasing opacity. Likewise, the
secondary maxima exhibited by the $JHK$ lightcurves $\sim$30 days after $B$ maximum
\citep{elias81,elias85,jha99,hernandez00,krisciunas00} are
due to the increasing release of energy through lower-optical depth IR
transitions. The prominent post-maximum emission features displayed by
SN~1999ee in the $H$ and $K$ bands lend support to this picture.


\subsection{SN~1999ex}
\label{sn99ex.sec}

SNe are classified according to their spectral features near maximum light \citep{filippenko97}.
Type II SNe are characterized by prominent hydrogen features and are believed to arise from
the core collapse of massive ($>$ 8-10 M$_{\odot}$) stars.  The common feature in all Type I SNe
is the lack of hydrogen spectral lines. The strong Si II $\lambda$6355 is the defining feature
of SNe Ia. Their occurrence in all type of stellar environments has led to the general consensus
that the progenitors of SNe~Ia are white dwarfs that undergo thermonuclear disruption 
after a period of mass transfer from binary companions. The proximity of SNe~Ib and Ic
to massive star formation regions \citep{vandyk96}, on the other hand, provides evidence that
these objects result from the core collapse of massive stars.  The lack of hydrogen
in their spectra is attributed to the loss of their outer hydrogen envelopes by stellar
winds or mass transfer to binary companions. There is growing evidence for a close physical
connection between SNe~II and SNe~Ib from the observations of SNe~1987K, 1993J, and 1996cb, which
evolved from SNe~II at early epochs to SNe~Ib at later times \citep{filippenko88,filippenko93,qiu99}.
The distinguishing difference between Type Ib and Ic SNe are the {\it optical} He I lines,
which are conspicuous in SNe Ib and weak or absent in SNe Ic. It is still a matter of debate
whether SNe~Ic constitute an extreme case of SNe~Ib or a completely separate
class of objects. The apparent absence of He in the spectra of SNe~Ic motivated the idea
that the progenitors of these objects could be nearly bare C+O cores of massive stars \citep{wheeler90}
that loose most of their helium by binary transfer or strong stellar winds. However, \citet{woosley95}
showed that helium stars could be the progenitors of both SNe~Ib and SNe~Ic.
They showed that the presence or absence of He I lines in the spectrum
is determined primarily by the amount of mixing, not the amount of
helium present -- greater mixing of radioactive material increases the
excitation of helium leading to stronger lines.

Significant effort has been put over recent years into determining the presence of helium
in the spectra of SNe~Ic in order to better understand the nature of these objects.
A detailed inspection of the spectra of the Type Ic SN~1994I
led \citet{filippenko95} to conclude that weak He I lines were probably present in the
optical region and that He I $\lambda$10830 was very prominent, although its Doppler
shift implied an unusually high expansion velocity $\sim$16,600 km~s$^{-1}$.
\citet{clocchiatti96} confirmed these observations and found evidence that He I $\lambda$5876
was also present in SN~1994I with a blueshift of $\sim$17,800 km~s$^{-1}$. They also reported 
high velocity He I $\lambda$5876 in the spectra of the three best-observed Type Ic SNe (1983V, 1987M, and 1988L),
which led them to conclude that most, and probably all, SNe~Ic show evidence for optical He I lines.
This conclusion, however, was recently questioned by \citet{millard99} and \citet{baron99} by means of
spectral synthesis models which showed that the IR feature near 10250 \AA~could
be accounted with lines of C I and Si I. Moreover, \citet{baron99} argued that
the feature attributed to He I $\lambda$5876 in the spectrum of SN~1994I could 
be a blend of other species.  Evidently, there is no consensus yet about the presence
or absence of He in the spectra of SNe~Ic.

Recently \citet{matheson01} compiled and analyzed a large collection of spectra of SNe~Ib and Ic.
This study showed no compelling evidence for He in the spectra of SNe~Ic and no gradual transition
from the Ib to the Ic class, which supported the idea that these objects are produced by different
progenitors.

\begin{figure}
\figurenum{10}
\epsscale{1.0}
\plotone{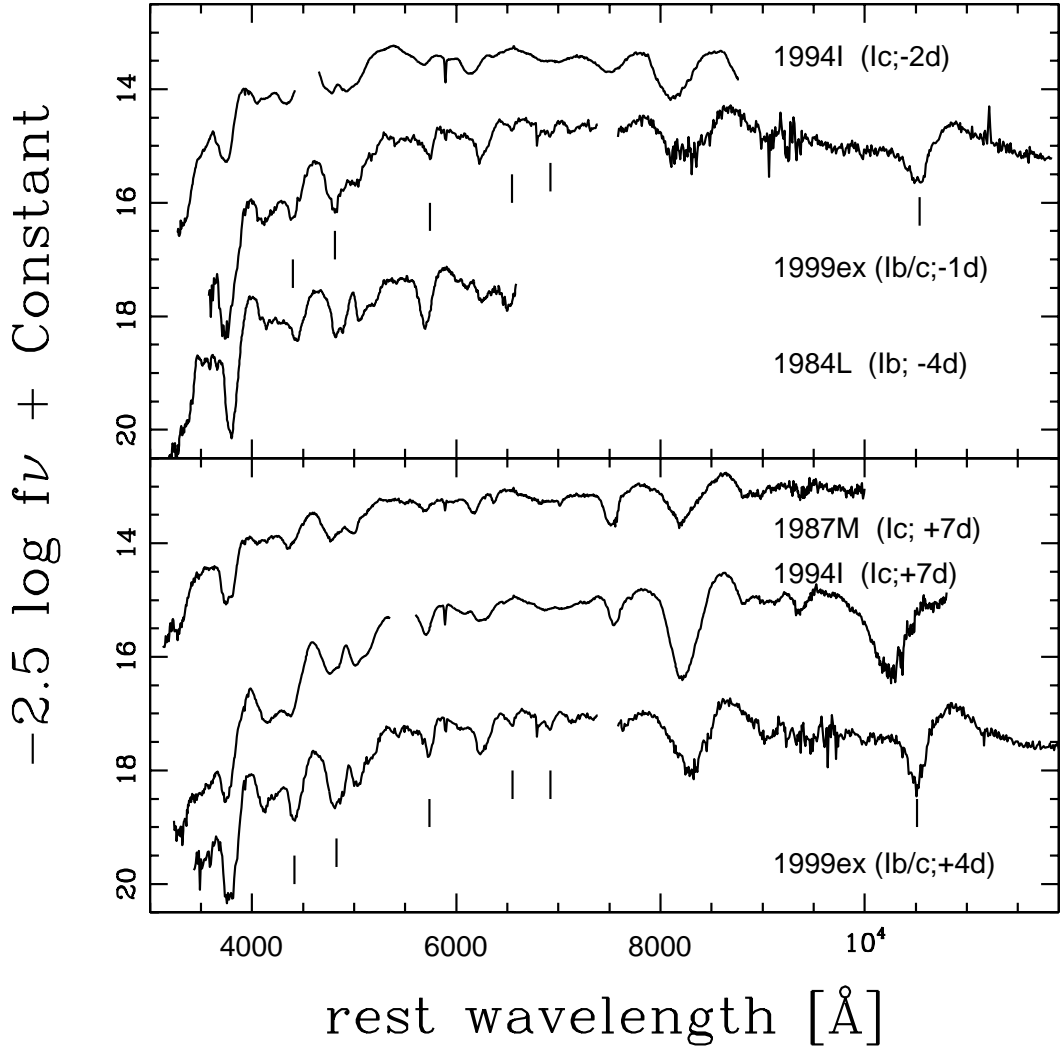}      
\caption{Comparison of near-maximum spectra of the Type Ib/c SN~1999ex with the prototype
of the Ib class SN~1984L, and the Type Ic SNe~1994I and 1987M.
Tick marks indicate the He I lines in the SN~1999ex spectra.
The strengths of the He lines gradually increase from the
Type Ic to the Ib SN, and SNe~1999ex appears to bridge the separation
between these two subclasses.
\label{sn99ex.opt.comp1.fig}}
\end{figure}

\begin{figure}
\figurenum{11}
\epsscale{1.0}
\plotone{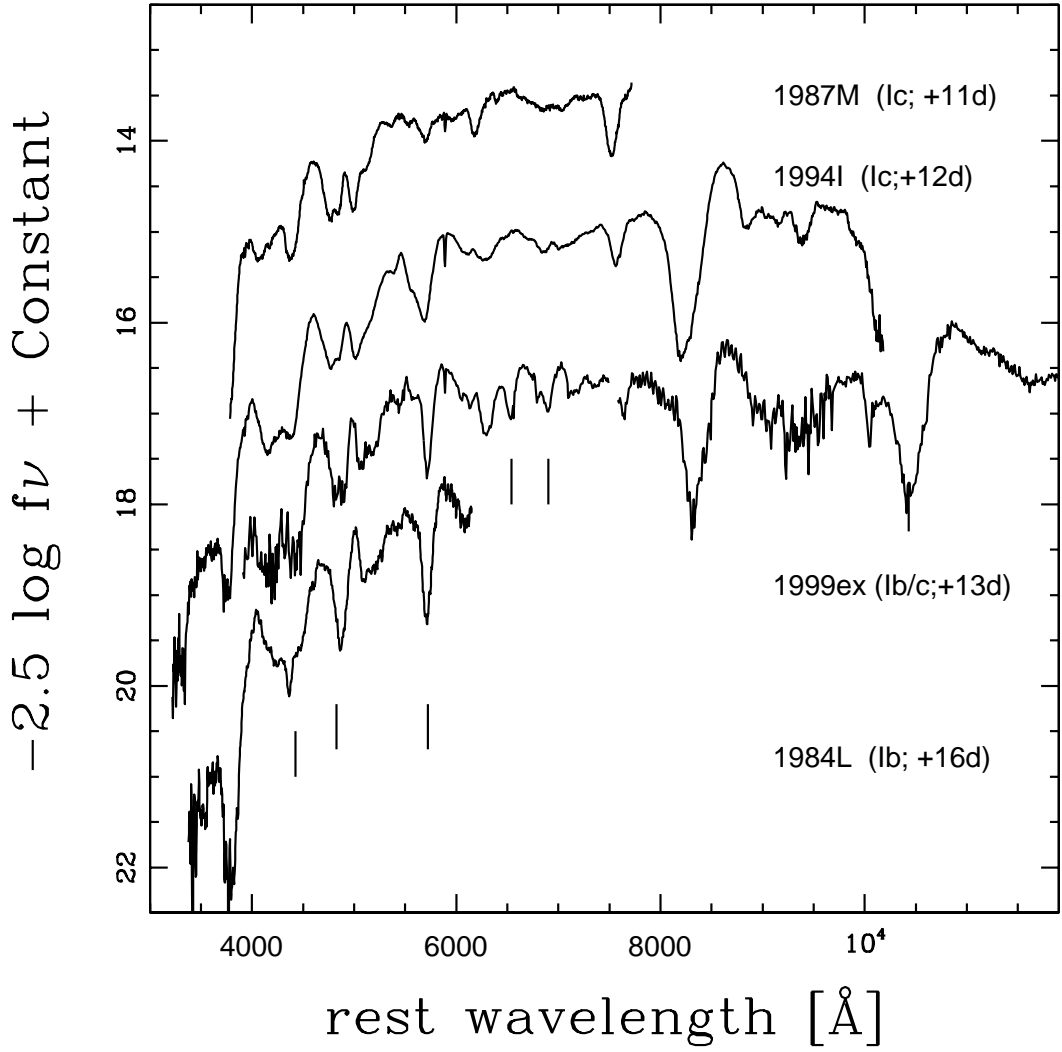}     
\caption{Same comparison as Figure \ref{sn99ex.opt.comp1.fig}, but
for spectra taken two weeks past maximum.
\label{sn99ex.opt.comp2.fig}}
\end{figure}

Our optical spectra of SN~1999ex show evidence for He I absorptions of moderate strength
in the optical region, thus suggesting the existence of an intermediate Ib/c case.
To illustrate this point in Figures \ref{sn99ex.opt.comp1.fig} and \ref{sn99ex.opt.comp2.fig}
we show a comparison between SN~1999ex, the best-observed Type Ic SN~1994I,
and the prototypes of the Ib (SN~1984L) and Ic class (SN~1987M). 
The near-maximum spectra (top panel of Figure \ref{sn99ex.opt.comp1.fig}) reveal a
gradual increase in the strength of all He lines (indicated with tick marks) from
the Ic to the Ib class. In the bottom panel of Figure \ref{sn99ex.opt.comp1.fig}
(one week past maximum) it is possible to see that, even though the 
spectra of the Type Ic SNe~1994I and 1987M are quite similar, SN~1994I shows
deeper troughs at the wavelengths of the He I lines, especially at 4471, 4921, and 5876 \AA.
The spectrum of SN~1994I obtained two weeks past maximum (Figure \ref{sn99ex.opt.comp2.fig})
is particularly interesting as it provides good evidence that the He I $\lambda$5876 line
was indeed present in the spectrum of SN~1994I with a higher expansion velocity than
the Na I D $\lambda$5893 doublet. This was also noted by \citet{filippenko95} who quoted
an expansion velocity of $\sim$16,600 km~s$^{-1}$ for the He line (see also \citet{clocchiatti96}
for a detailed discussion). Overall, Figures \ref{sn99ex.opt.comp1.fig}-\ref{sn99ex.opt.comp2.fig}
reveal that there is a smooth spectroscopic sequence ranging from the He deficient Type Ic
SNe~1987M and 1994I, the Type Ib/c SN~1999ex, and the Type Ib SN~1984L that is characterized by
strong optical He lines.  We conclude therefore that SN~1999ex provides first evidence
for a link between the Ib and Ic classes and that there is a continuous sequence of SNe Ib and Ic
objects. 

Special attention must be paid to the IR feature near 11,000 \AA~which was very prominent
in SNe~1999ex and 1994I. As mentioned above the definitive identity of this feature in
SN~1994I is still uncertain: it could be accounted with lines of C I and Si I
\citep{millard99,baron99} or with high velocity He I $\lambda$10830 \citep{filippenko95,clocchiatti96}.
If the feature in SN~1999ex was He I $\lambda$10830 it would imply
a moderate velocity of 6,000-8,000 km~s$^{-1}$, which matches very well 
the velocities derived from the Fe, Na and Si lines. The presence of the
He I $\lambda$20581 with the same expansion velocity (see Figure \ref{sn99ex.ir.fig}) suggests that 
the IR feature in SN~1999ex was indeed due to He I $\lambda$10830. Altogether
our IR spectra of SN~1999ex provide unambiguous proof that He was present
in the atmosphere of this intermediate Ib/c object. A detailed atmosphere
model could be very useful at placing limits on the He mass in the
ejecta of SN~1999ex and constraining the nature of its progenitor.
As suggested by \citet{clocchiatti96} the presence of highly blueshifted
He I $\lambda$5876 in SN~1994I suggests that the IR feature was also due to He I $\lambda$10830.
If so, the difference in He velocities between SNe~1994I and 1999ex might prove an interesting clue to the
underlying physics of the atmospheres of this class of objects.
Further optical/IR spectroscopy and photometry will lead us
to better understanding the atmospheres of SNe~Ib and Ic.

\begin{figure}
\figurenum{12}
\epsscale{1.0}
\plotone{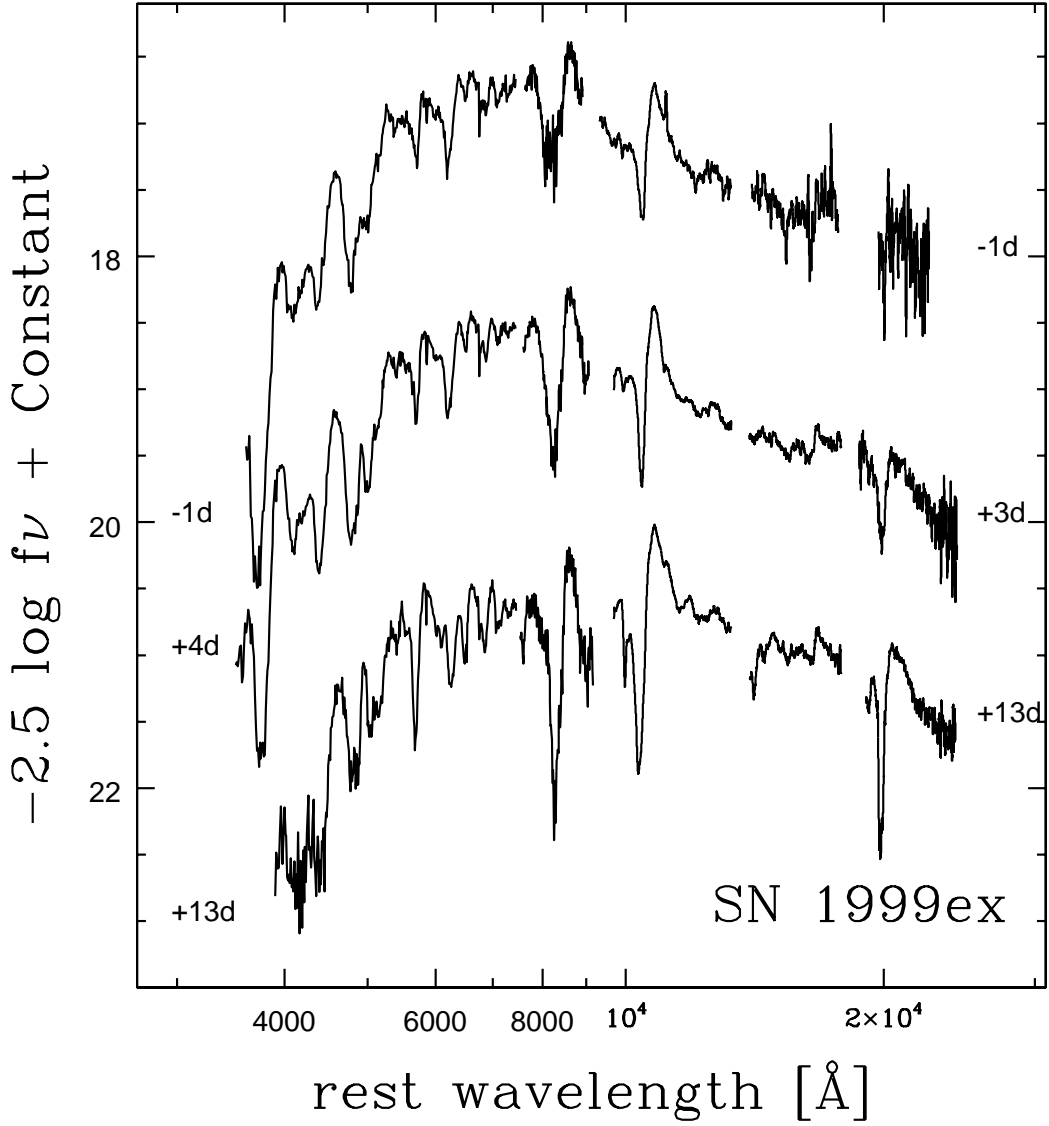} 
\caption{Combined optical and IR spectra of SN~1999ex
in AB magnitudes. Days since $B$ maximum are indicated next to each spectrum.
\label{sn99ex.fig}} 
\end{figure}

Finally, given that the IR spectra were obtained within one
day from the optical spectra, we were able to combine
these observations. Figure \ref{sn99ex.fig} shows the resulting
spectra and the excellent agreement between the optical and IR fluxes.
Note the prominent He I $\lambda$$\lambda$10830,20581 features.

\subsection{Maximum-light optical/IR spectra of SNe}

\begin{figure}
\figurenum{13}
\epsscale{1.0}
\plotone{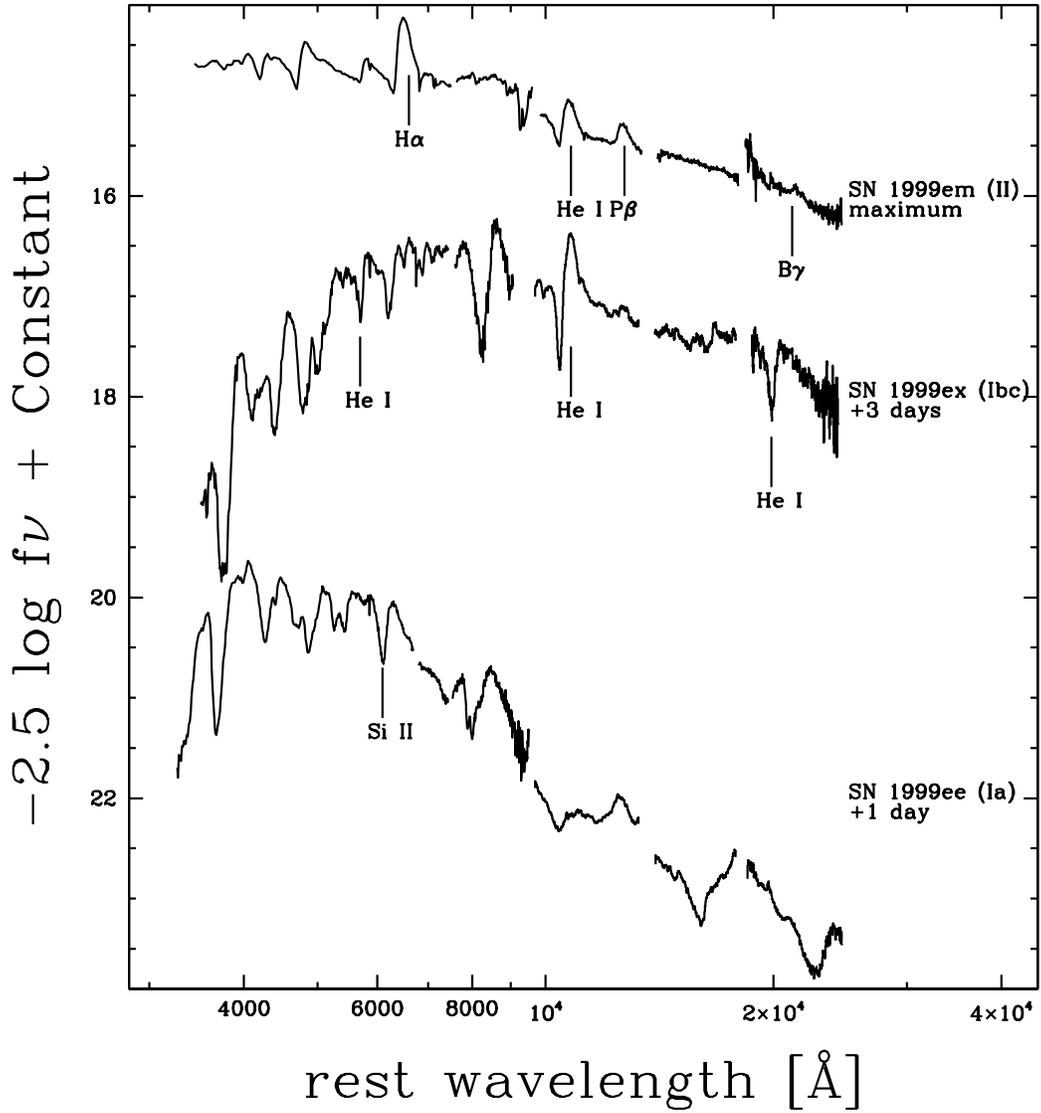} 
\caption{Combined optical and IR maximum-light spectra of the Type II SN~1999em, the Type Ib/c SN~1999ex,
and the Type Ia SN~1999ee.
\label{allsne.fig}}
\end{figure}

In the course of the SOIRS program we obtained high-quality
spectroscopy of the Type II SN~1999em \citep{hamuy01}.
Figure \ref{allsne.fig} compares maximum-light spectra of the Type
II SN~1999em, the Type Ib/c SN~1999ex, and the Type Ia SN~1999ee.
This figure permits one to compare the different characters
of the main classes of SNe, all the way from 3,000 to 25,000 \AA.
The Type II is distinguished by prominent hydrogen Balmer
and Paschen lines. It also shows the P-Cygni profile of
the He I $\lambda$10830 transition.
The Type Ib/c is characterized by the strong
He I $\lambda\lambda$10830,20581 lines in the IR, the He I $\lambda$5876
line in the optical, and other lines in the optical due to
singly ionized Ca, Fe, Mg, Si and neutral Na and O. It is
interesting to note that, while both the Type II and the Type Ib/c show
strong He I $\lambda$10830, only the Type Ib/c shows a significant He I $\lambda$20581
feature.  Finally, the Type Ia at the bottom of this figure is characterized
by the strong Si II $\lambda$6355 and other intermediate-mass
elements, and the absence of hydrogen and helium.

\section{Conclusions}

The observations obtained for SN~1999ee constitute the most complete
spectral and temporal coverage ever achieved for a SN Ia. Its Branch-normal
character makes it an ideal reference for comparative studies of SNe~Ia.

Before maximum light SN~1999ee displayed a normal spectrum with a strong
Si II $\lambda$6355 absorption, thus showing that not all slow-declining SNe
are spectroscopically peculiar at these evolutionary phases. We conclude
that the photometric properties of luminous SNe~Ia cannot be used to predict
spectroscopic peculiarities.

From a comparison of the IR spectra of SN~1999ee and other SNe~Ia that encompass
a wide range in decline rates we find that there is a remarkable homogeneity among
the Branch-normal SNe Ia during their first 60 days of evolution. Although the slow-decliner
luminous SN~2000cx showed a premaximum featureless IR spectrum similar to that of
normal SNe, the Mg II $\lambda10926$ line was characterized by a high expansion velocity.
The fast-decliner subluminous SN~1999by was noticeably different than the other SNe
at all epochs. This study reveals that the spectroscopic peculiarities displayed by
SN~1991bg-like objects at optical wavelengths are also present in the IR.

The fortunate occurrence of SN~1999ex within three weeks and in the same galaxy
that hosted SN~1999ee permitted us to obtain optical and IR spectroscopy of
a Ib/c event. SN~1999ex was characterized for the lack of hydrogen lines,
weak optical He I lines, and strong He I $\lambda$10830,20581, thus
providing an example of an intermediate case between pure Ib and Ic SNe.
We conclude therefore that SN~1999ex provides first clear evidence
for a link between the Ib and Ic classes and that there is a continuous
spectroscopic sequence ranging from the He deficient SNe~Ic to the SNe~Ib
which are characterized by strong optical He I lines. 

\acknowledgments

\noindent

M. H. is very grateful to Las Campanas and Cerro Cal\'an observatories
for allocating an office and providing generous operational support to
the SOIRS program during 1999-2000.
M. H. and J. M. thank the ESO, CTIO, and Las Campanas visitor support
staffs for their assistance in the course of our observing runs, 
and  the Director General of ESO for allocating
Director's discretionary telescope time to this project.
We are very grateful to A. Filippenko, P. H\"oflich, S. Jha, D. Leonard, W. Li, P. Meikle,
R. Rudy, and J. Spyromilio, for making us available their spectra
of SNe~1984L, 1987M, 1994D, 1994I, 1999by, 1998bu, and 2000cx.
Support for this work was provided by NASA through Hubble Fellowship grant HST-HF-01139.01-A
awarded by the Space Telescope Science Institute, which is operated by the Association
of Universities for Research in Astronomy, Inc., for NASA, under contract NAS 5-26555.
J.M. acknowledges support from FONDECYT grant 1980172.
This research has made use of the NASA/IPAC Extragalactic Database (NED), which is operated by the
Jet Propulsion Laboratory, California Institute of Technology, under
contract with the National Aeronautics and Space Administration.
This research has made use of the SIMBAD database, operated at
CDS, Strasbourg, France.

\begin{deluxetable} {lccccccccccc}
\tabletypesize{\scriptsize}
\rotate
\tablecolumns{12}
\tablenum{1}
\tablewidth{0pc}
\tablecaption{Journal of the Spectroscopic Observations}
\tablehead{
\colhead{UT Date} & 
\colhead{Julian Date} &
\colhead{Observ-} &
\colhead{Telescope} &
\colhead{Instr-} &
\colhead{Detector} &
\colhead{Grating/} &
\colhead{Order} &
\colhead{Wavelength} &
\colhead{Dispersion} &
\colhead{Weather} &
\colhead{Observer(s)} \\
\colhead{} &
\colhead{-2451000} &
\colhead{atory} &
\colhead{} &
\colhead{ument} &
\colhead{} &
\colhead{Grism} &
\colhead{} &
\colhead{($\mu m$)} &
\colhead{(\AA/pix)} & 
\colhead{} &
\colhead{}}
\startdata

1999 Oct 09 & 460.53 & Paranal      & VLT/Antu    & ISAAC     &Rockwell1K     & \nodata   &5,4,3,2 & 0.98-2.50 &2.9,3.6,4.7,7.1&  Clear   & Hamuy,Lidman \\
1999 Oct 09 & 460.56 & Tololo       & 4-m         & R-C Spec. &Loral3Kx1K     & KPGL-2    &  1     & 0.33-0.87 & 1.9           &  Cirrus  & Maza \\
1999 Oct 11 & 462.52 & La Silla     & NTT         & EMMI      &TEK1024+2048   & 5,13      &  1     & 0.33-1.01 & 2.7,3.5       &  Clear   & Hamuy,Doublier \\
1999 Oct 16 & 467.56 & Campanas     & 2.5-m       & WFCCD     &TEK2048        & Blue      &\nodata & 0.36-0.92 & 3.0           &  Clouds? & Phillips  \\
1999 Oct 18 & 469.49 & La Silla     & NTT         & SOFI      &Rockwell1K     &  Red      &\nodata & 1.50-2.53 & 10.2          &  Clear   & Doublier,Maza \\
1999 Oct 18 & 469.60 & La Silla     & NTT         & EMMI      &TEK1024+2048   & 5,13      &  1     & 0.33-1.00 & 2.7,3.5       &  Clear   & Maza \\
1999 Oct 19 & 470.50 & Paranal      & VLT/Antu    & ISAAC     &Rockwell1K     & \nodata   &5,4,3,2 & 0.98-2.50 &2.9,3.6,4.7,7.1&  Clear   & Hamuy,Cuby,Petr \\
1999 Oct 20 & 471.56 & La Silla     & D1.5-m      & DFOSC     &Loral2Kx2K     & 3,5       &  1     & 0.33-0.98 & 2.3,3.1       &  \nodata & Pompei \\
1999 Oct 22 & 473.61 & La Silla     & NTT         & SOFI      &Rockwell1K     &Blue,Red   &\nodata & 0.95-2.47 & 7.0,10.2      &  Clear   & Hainaut,Lefloc'h \\
1999 Oct 25 & 476.51 & La Silla     & D1.5-m      & DFOSC     &Loral2Kx2K     & 3,5       &  1     & 0.33-0.98 & 2.3,3.1       &  Clear   & Augusteyn \\
1999 Oct 26 & 477.48 & La Silla     & NTT         & SOFI      &Rockwell1K     &Blue,Red   &\nodata & 0.95-2.46 & 7.0,10.2      &  \nodata & Hainaut,Vanzi \\
1999 Oct 27 & 478.52 & Tololo       & 1.5-m       & R-C Spec. &Loral1.2Kx0.8K &  58       &  2     & 0.37-0.50 & 1.1           &  Clear   & Olsen \\
1999 Oct 29 & 480.51 & La Silla     & D1.5-m      & DFOSC     &Loral2Kx2K     & 3,5       &  1     & 0.34-0.98 & 2.3,3.1       &  \nodata & Augusteyn \\
1999 Nov 02 & 484.49 & Paranal      & VLT/Antu    & ISAAC     &Rockwell1K     & \nodata   &5,4,3,2 & 0.98-2.50 &2.9,3.6,4.7,7.1&  Clear   & Hamuy,Lidman,Petr \\
1999 Nov 03 & 485.62 & La Silla     & NTT         & EMMI      &TEK1024+2048   & 5,13      &  1     & 0.33-1.01 & 2.7,3.5       &  \nodata & Maza \\
1999 Nov 06 & 488.52 & La Silla     & NTT         & SOFI      &Rockwell1K     & Blue      &\nodata & 0.94-1.65 & 7.0           &  Clear   & Brillant \\
1999 Nov 06 & 488.53 & La Silla     & D1.5-m      & DFOSC     &Loral2Kx2K     & 3,5       &  1     & 0.34-0.98 & 2.3,3.1       &  Clear?  & Augusteyn \\
1999 Nov 09 & 491.50 & La Silla     & NTT         & SOFI      &Rockwell1K     &Blue,Red   &\nodata & 0.95-2.51 & 7.0,10.2      &  Clear   & Hamuy,Brillant \\
1999 Nov 09 & 491.62 & La Silla     & NTT         & EMMI      &TEK1024+2048   & 5,13      &  1     & 0.34-1.00 & 2.7,3.5       &  Clear   & Hamuy,Brillant \\
1999 Nov 14 & 496.56 & La Silla     & NTT         & SOFI      &Rockwell1K     &Blue,Red   &\nodata & 0.95-2.46 & 7.0,10.2      &  Clear   & Hamuy,Doublier \\
1999 Nov 14 & 496.62 & La Silla     & NTT         & EMMI      &TEK1024+2048   & 5,13      &  1     & 0.34-1.00 & 2.7,3.5       &  Clear   & Hamuy,Doublier  \\
1999 Nov 18 & 500.53 & Paranal      & VLT/Antu    & ISAAC     &Rockwell1K     & \nodata   &5,4,3,2 & 0.98-2.50 &2.9,3.6,4.7,7.1&  Clear   & Serv. Obs. \\    
1999 Nov 19 & 501.58 & La Silla     & NTT         & EMMI      &TEK1024+2048   & 5,13      &  1     & 0.34-1.00 & 2.7,3.5       &  Clear   & Hamuy,Doublier \\ 
1999 Nov 28 & 510.52 & Paranal      & VLT/Antu    & ISAAC     &Rockwell1K     & \nodata   &5,4,3,2 & 0.98-2.50 &2.9,3.6,4.7,7.1&  Clear   & Hamuy,Lidman,Chadid \\
1999 Nov 28 & 510.63 & La Silla     & D1.5-m      & DFOSC     &Loral2Kx2K     & 3,5       &  1     & 0.40-0.98 & 2.3,3.1       &  \nodata & Pinfield \\

\enddata
\end{deluxetable}
\clearpage

\end{document}